\documentclass[a4paper,11pt]{article}
\usepackage{jinstpub}
\makeatletter
\gdef\@fpheader{}
\makeatother

\usepackage{float}
\usepackage{subcaption}
\usepackage{appendix}
\usepackage{xcolor}

\title{\boldmath Performance Benchmarks for 2-View and 3-View Fiber-Projection Fine-Grained Particle Detectors}

\author[a]{Haohui Che}
\author[b,1]{and Guang Yang\note{Corresponding author.}}

\affiliation[a]{Washington University in St. Louis,\\St. Louis, MO 63130, USA}
\affiliation[b]{Brookhaven National Laboratory,\\Upton, NY 11973, USA}

\emailAdd{gyang1@bnl.gov}

\abstract{
\noindent Fine-grained scintillator detectors are critical for precision measurements in nuclear and particle physics, where accurate reconstruction of interaction vertices and secondary particle directions enables separation of signal from background events. A well-known design choice is the fiber readout geometry: traditional 2-View systems use orthogonal X and Y fibers, while next-generation 3-View designs add a third Z-fiber layer that provides unambiguous 3D voxel identification. The 2-View approach suffers from combinatorial ``ghost hits'', which are false 3D candidates arising from fiber projection ambiguities, that degrade reconstruction performance in high-multiplicity events. This paper presents comprehensive simulation benchmarks quantifying the performance difference between 2-View and 3-View geometries across key metrics. We find that the 3-View geometry reduces ghost hits by 30 to 90\% depending on event topology, provides robust vertex resolution across complex topologies, and maintains superior angular resolution for shower direction reconstruction. These benchmarks inform the design optimization of future detectors and provide quantitative guidance for reconstruction algorithm development across a broad range of experiments including neutrino physics, rare kaon/pion decays, and collider calorimetry.
}


\begin{document}

\maketitle

%==============================================================================
\section{Introduction}
\label{sec:intro}
%==============================================================================

Fine-grained scintillator detectors have become essential tools across multiple frontiers of nuclear and particle physics. Their combination of tracking capability and calorimetric response makes them particularly suitable for experiments requiring precise reconstruction of complex final states. This paper presents comprehensive simulation benchmarks comparing two fundamental readout geometries: the traditional 2-View and the next-generation 3-View approach. We quantify the performance gains in ghost hit suppression, angular resolution, vertex resolution, and energy-dependent particle reconstruction.

We focus on the energy range 0.1 to 10~GeV, most relevant for accelerator-based neutrino experiments, rare meson decay searches, and test-beam characterization of calorimeter prototypes. The dominant event topologies in this regime include quasi-elastic scattering (single charged track), resonant pion production (lepton plus pions), and deep inelastic scattering (high-multiplicity hadronic final states), as well as electromagnetic showers from $\pi^0 \to \gamma\gamma$ decay. These cover hit multiplicities from $\mathcal{O}(10)$ to $\mathcal{O}(1000)$ and a wide range of angular distributions, making them well suited for evaluating the readout geometry.

Throughout this paper, \textit{vertex resolution} refers to the 68th-percentile error in the reconstructed spatial position of the interaction point (or shower origin), measured in centimeters. \textit{Angular resolution} refers to the 68th-percentile angular deviation between the true and reconstructed shower or track direction, measured in degrees.

\subsection{Physics Motivation}

The demand for fine-grained 3D hit reconstruction spans a remarkably diverse range of physics programs. Firstly, in the neutrino oscillation experiments such as the ongoing T2K \cite{T2K_Oscillation}, NOvA \cite{NOvA_Oscillation} experiments, require near detectors that precisely reconstruct the interaction vertex and identify outgoing particle multiplicities to constrain neutrino-nucleus cross sections \cite{MINERvA}. Traditional 2-View geometry has been successfully employed in operating experiments such as the T2K Fine-Grained Detectors \cite{T2K_FGD} and the NOvA detector \cite{NOvA}, which demonstrated few-centimeter vertex resolution. The upgraded SuperFGD extends this with 3-View readout \cite{Abe:2019whr, Blondel:2018tft}. Future experiments including DUNE \cite{DUNE_TDR} and Hyper-Kamiokande \cite{HyperK_TDR} are considering similar concepts.
In addition, experiments searching for rare kaon decays such as NA62 ($K^+ \to \pi^+ \nu \bar{\nu}$) \cite{NA62_Detector, NA62_Result} and KOTO ($K_L \to \pi^0 \nu \bar{\nu}$) \cite{KOTO_Detector} require hermetic photon vetoes to reject backgrounds from $\pi^0 \to \gamma\gamma$. The PIONEER experiment \cite{PIONEER, PiENu} probes lepton universality through precision $\pi^+ \to e^+ \nu$ measurements, demanding excellent tracking in the decay region. Furthermore, Particle Flow Calorimetry at future $e^+e^-$ colliders (ILC, CLIC, FCC-ee) requires unprecedented spatial granularity to separate nearby showers \cite{CALICE_Review}. The CMS HGCAL \cite{CMS_HGCAL} and EIC detectors \cite{EIC_CDR, ePIC_Detector} also employ highly granular scintillator sampling sections.
Additionally, fine-grained scintillators also find use in dark matter veto systems, muon tomography, medical beam monitoring, and cosmic ray physics. A systematic understanding of the advantage of the 3D-projection readout over the 2D-projection readout through optical fiber is beneficial to the future particle detector design and optimization.

\subsection{3-View Detector Concept}

\begin{figure}[H]
    \centering
    \includegraphics[width=0.85\linewidth]{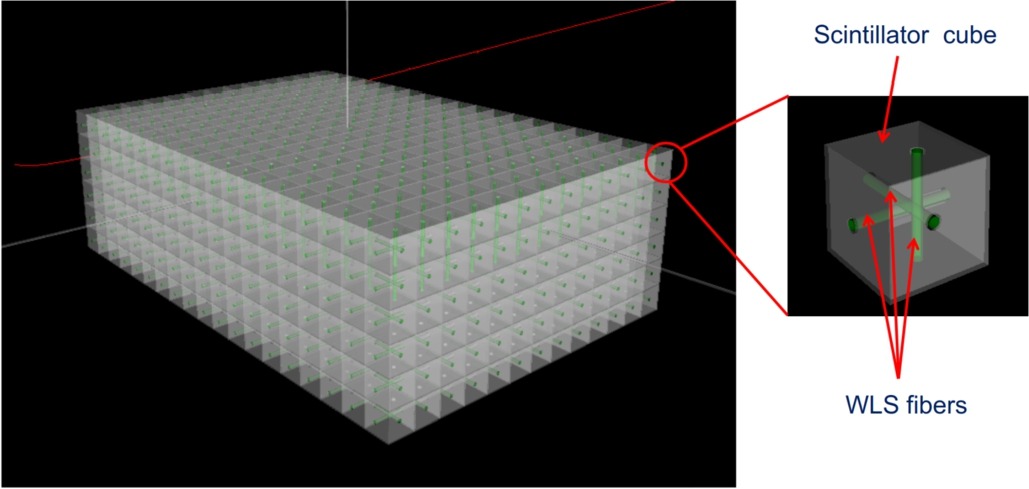}
    \caption{The 3-View scintillator detector concept. Each cubic voxel is read out by three orthogonal wavelength-shifting fibers (X, Y, Z), with MPPCs at the fiber ends. The highlighted cube illustrates the triple-fiber readout geometry. This concept was developed for the T2K ND280 Upgrade SuperFGD \cite{Blondel:2018tft, Abe:2019whr}.}
    \label{fig:detector_concept}
\end{figure}

The fundamental architecture of a 3-View (3V) scintillator detector is illustrated in Figure~\ref{fig:detector_concept}. The active volume consists of an array of optically isolated scintillator cubes (voxels), each penetrated by three orthogonal wavelength-shifting (WLS) fibers running along the X, Y, and Z directions. When a charged particle traverses a cube and deposits energy via ionization, scintillation light is produced isotropically within the cube. A fraction of this light is captured by each of the three WLS fibers, wavelength-shifted, and guided to Multi-Pixel Photon Counters (MPPCs) located at the fiber ends.

In a traditional 2-View (2V) geometry, only X and Y fibers are employed. A hit at position $(x, y, z)$ lights the X-fiber passing through $(y, z)$ and the Y-fiber passing through $(x, z)$. Reconstruction identifies candidate hit positions by finding all combinations where both an X-fiber and a Y-fiber are lit at the same $z$-layer.

\subsection{Ghost Hit Problem}

The 2-View geometry suffers from a fundamental combinatorial ambiguity. Consider two true hits at positions $(x_1, y_1, z)$ and $(x_2, y_2, z)$ in the same $z$-layer. The detector records four lit fibers: X-fibers at $(y_1, z)$ and $(y_2, z)$, and Y-fibers at $(x_1, z)$ and $(x_2, z)$. The reconstruction algorithm evaluates all fiber crossings, yielding four candidates: the two true hits $(x_1, y_1)$ and $(x_2, y_2)$, plus two ``ghost'' candidates $(x_1, y_2)$ and $(x_2, y_1)$ that do not correspond to real energy deposits.

In general, $N$ true hits in a single $z$-layer produce $N^2$ candidates, of which only $N$ are genuine and $N^2 - N$ are ghosts. This quadratic scaling creates severe problems in high-multiplicity events. Ghost contamination:
Ghost contamination degrades vertex resolution by biasing the reconstructed position toward ghost centroids and corrupts angular reconstruction by introducing spurious hit directions. Furthermore, it inflates apparent energy deposits, biasing calorimetric measurements; in practice, computing energy deposits from 1D fiber projections rather than 3D candidates can mitigate this effect in 2-View systems. Ghost contamination also complicates pattern recognition by creating false track and shower candidates.

The 3-View geometry introduces a third constraint through the Z-fiber. A true hit at $(x, y, z)$ now lights three fibers: X-fiber at $(y, z)$, Y-fiber at $(x, z)$, and Z-fiber at $(x, y)$. A candidate hit is accepted only if \emph{all three} corresponding fibers register signals.

For the example above with two true hits, the Z-fibers lit are only those at $(x_1, y_1)$ and $(x_2, y_2)$. The ghost candidates at $(x_1, y_2)$ and $(x_2, y_1)$ fail the Z-fiber requirement and are rejected. The 3-View geometry thus provides natural ghost suppression through topological constraints, without requiring energy matching or timing information.

However, the 3-View geometry does not eliminate \emph{all} ghosts. Certain symmetric hit configurations, such as four hits arranged on a rectangular prism, create fiber coincidences that satisfy the triple constraint at ghost positions as well. These irreducible ghost hits represent a fundamental limit of the light-projection-based approach.

\subsection{Paper Purpose and Structure}

Despite the conceptual clarity of the 3-View advantage, quantitative benchmarks are essential for detector optimization. This work addresses several key questions such as the ghost reduction factor as a function of event topology and hit multiplicity, 3-View improvement on angular resolution for shower direction reconstruction, vertex resolution gain that achieved for single and multi-shower events, and the improvement scaling with particle energy and scintillator granularity.
This paper addresses these questions through comprehensive Monte Carlo simulation, providing quantitative guidance for detector design decisions. The results are applicable to a broad range of experiments employing fine-grained scintillator technology.

The paper is organized as follows. Section~\ref{sec:method} describes the simulation framework, including the detector model, particle generation, optical assumptions, and reconstruction algorithms. Section~\ref{sec:ghosts} quantifies ghost hit suppression across different event topologies, from track-like to shower-like configurations, and presents analytical derivations of the expected ghost counts. Section~\ref{sec:angular} evaluates single-shower angular resolution as a function of shower parameters and particle direction. Sections~\ref{sec:single_vertex} and \ref{sec:two_cluster} present vertex resolution studies for single-shower and two-shower topologies, respectively. Section~\ref{sec:energy} examines energy-dependent performance for realistic muon and electron simulations spanning 100~MeV to 10~GeV. Finally, Section~\ref{sec:conclusion} summarizes key findings and discusses implications for future detector designs. Appendix~\ref{app:fine_pitch} replicates all studies at finer pitch (0.5 cm) to demonstrate performance scaling.

%==============================================================================
\section{Simulation Framework}
\label{sec:method}
%==============================================================================

\subsection{Detector Model}

We model a generic fine-grained scintillator detector as a cubic voxelized volume with baseline pitch $p = 1.0$ cm. The detector spans approximately 1 m$^3$ in the simulation, though results and conclusions are largely independent of detector size for the topologies studied. All main results use the 1.0 cm baseline. Appendix~\ref{app:fine_pitch} presents a complete replication of all studies at finer pitch ($p = 0.5$ cm) to validate performance scaling and demonstrate that further improvements are achievable with even smaller voxel sizes.

\subsection{Optical Simulation Assumptions}

In this study, we employ an idealized optical model where scintillation light is perfectly confined within each cube-shaped voxel and is collected only by the three orthogonal wavelength-shifting (WLS) fibers penetrating that voxel. This ``perfect confinement'' approximation does not account for optical crosstalk, i.e., the leakage of scintillation photons through the fiber holes into neighboring cubes.

Measurements with physical prototypes of the SuperFGD detector \cite{Blondel:2018tft, Abe:2019whr} have characterized this optical crosstalk. Light leakage through the $\sim$1 mm diameter fiber holes was measured to be approximately 3\% per face, with total crosstalk to nearest-neighbor cubes estimated at 5 to 10\% depending on the reflective coating quality. This crosstalk can either be treated as a systematic effect that degrades resolution, or, with appropriate calibration, that can be exploited as additional position information to improve hit localization beyond the voxel pitch limit.

The perfect confinement assumption in our simulation therefore represents a conservative scenario. In practice, crosstalk signals provide sub-voxel position information that can enhance angular and vertex resolution. Future simulations incorporating crosstalk modeling are expected to show modest improvements beyond the results presented here, particularly at coarser pitch values where the crosstalk fraction relative to the primary signal is smaller.

\subsection{Electronics and Noise Considerations}

Our simulation assumes noiseless hit detection, with which a voxel is marked as ``lit'' if and only if a charged particle trajectory passes through it. We do not include any  thermal noise, dark count rates of the Multi-Pixel Photon Counters (MPPCs/SiPMs), or readout electronics noise.

This idealization is justified for several reasons. Modern MPPCs achieve dark count rates of order $10^5$ to $10^6$ Hz at room temperature \cite{Acerbi:2019wpn}, which corresponds to $\lesssim 0.1$\% occupancy per voxel in a 100 ns readout window, which is negligible compared to the signal hit densities considered in this study. The SuperFGD detector employs customized MPPCs with photon detection efficiency $>$25\% and crosstalk probability $<$3\% \cite{Mineev:2018ekk}. Combined with front-end ASICs providing thresholds of 0.5 to 1 photoelectron equivalent, the false hit rate from electronics noise is suppressed to levels that do not significantly impact reconstruction.

Furthermore, the ghost hit phenomenon under study is a combinatorial effect arising from fiber projection ambiguities, not from noise hits. The addition of a small random noise floor would increase the total candidate count slightly but would not qualitatively change the 2-View vs. 3-View comparison, as noise affects both geometries equally.

For these reasons, we expect our noiseless assumption to accurately represent the physics of ghost hit formation and the relative performance gains of 3-View readout. Quantitative predictions for specific detector implementations should incorporate realistic noise models calibrated to prototype measurements.

\subsection{Particle Generation and Geometry}

\begin{figure}[H]
    \centering
    \includegraphics[width=0.5\linewidth]{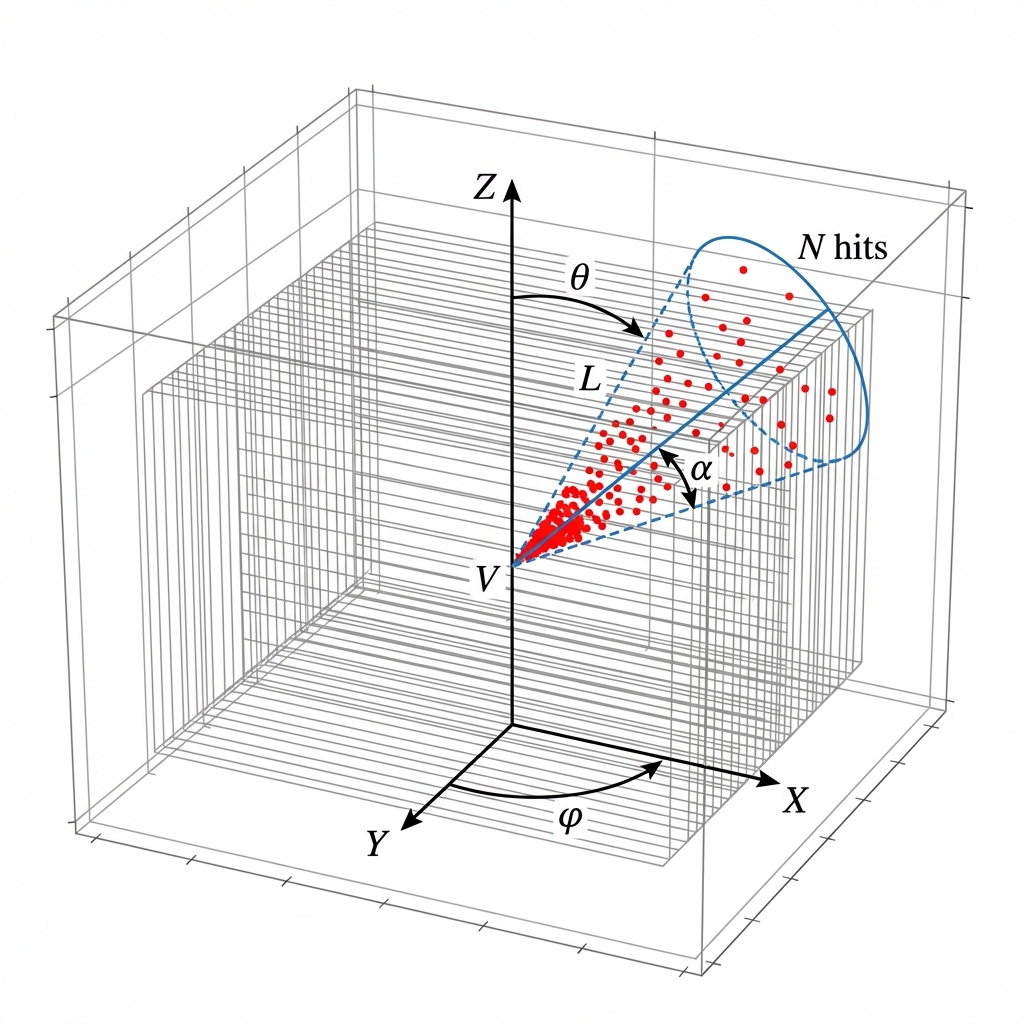}
    \caption{Geometry parameters for shower simulation. A cone-shaped shower originates from vertex $V$ with: $N$ hits distributed within the cone volume, length $L$ along the shower axis, polar angle $\theta$ from the Z-axis, azimuthal angle $\phi$ in the XY plane, and cone half-opening angle $\alpha$. Some of the 2V fiber grid is shown for reference.}
    \label{fig:cluster_geometry}
\end{figure}

The simulation generates particles with controlled geometric parameters to systematically explore the reconstruction phase space. Figure~\ref{fig:cluster_geometry} illustrates the key parameters for single-cluster and two-cluster topologies.

\medskip
\textbf{Single-Cluster Parameters:}
Single-cluster parameters include the number of hits $N$, where higher values could correspond to higher-energy particles. Starting from a toy model, we look into $N \in \{20, 50, 100\}$. The longitudinal extent $L$ varies between 10 to 40 cm, representing different particle penetrations or material densities. The cone opening angle $\alpha$ ranges from narrow, track-like topologies ($\alpha \lesssim 10^\circ$) to wide, shower-like deposits ($\alpha \gtrsim 20^\circ$), with specific values ranging from $10^\circ$ to more than $75^\circ$. Finally, the polar angle $\theta$ is varied across $0^\circ$ to $90^\circ$ in order to capture direction-dependent effects.

\medskip
\textbf{Two-Cluster Parameters (for vertex studies):}
For two-cluster vertex studies, we vary the separation angle between the two cluster axes from $10^\circ$ (nearly parallel) to $90^\circ$ (perpendicular). Each cluster is assigned a cone opening angle at a level of $20^\circ$ to $30^\circ$.

We generate particles as geometric objects, including tracks defined as line segments with specific start points, directions, and lengths. Showers are modeled as cones with hits uniformly distributed in azimuth and following a uniform radial profile. Additionally, full electromagnetic cascades are simulated, incorporating bremsstrahlung and pair production for energy-dependent studies in Section 7. In Section 4-6,
a voxel is considered ``lit'' if a particle's trajectory passes through it. For cone-shaped showers, hits are placed at random positions within the cone volume, then voxelized to the detector grid.

In Sections~\ref{sec:ghosts} through \ref{sec:two_cluster}, hits are distributed uniformly within the cone volume. Real showers have a denser core and sparser tails (roughly following a Moli\`ere profile), which would push the linearity $\lambda$ higher and shift results toward the track-like end of our parameter space. This tends to increase 2-View ghost counts (more fiber overlaps in the dense core) while the Z-fiber constraint in 3-View becomes even more effective for concentrated distributions. Our uniform-cone model is therefore conservative regarding the 3-View advantage.

\subsection{Fiber Readout denotation}

Let $S_X$, $S_Y$, $S_Z$ denote the sets of lit fibers in each view. For 2-View reconstruction, candidates are formed by matching fibers at the same $z$:
\begin{equation}
    C_{2V} = \bigcup_z \{ (x, y, z) \mid (y, z) \in S_X \wedge (x, z) \in S_Y \}
\end{equation}

For 3-View reconstruction, an additional constraint is applied:
\begin{equation}
    C_{3V} = \{ (x, y, z) \mid (y, z) \in S_X \wedge (x, z) \in S_Y \wedge (x, y) \in S_Z \}
\end{equation}

Ghost hits are defined as candidates that do not correspond to true energy deposits: $N_\text{ghost} = |C| - N_\text{true}$.

\subsection{Reconstruction Algorithms}

\textbf{Vertex finding}: We employ a Principal Component Analysis (PCA) axis extrapolation algorithm that achieves resolution scaling of $\sigma \sim p/\sqrt{N}$, where $p$ is the voxel pitch and $N$ is the number of hits. The algorithm proceeds as follows:
\begin{enumerate}
    \item \textbf{Compute PCA}: Find the principal axis of the hit distribution via eigendecomposition of the covariance matrix.
    \item \textbf{Project hits}: Project all hits onto the principal axis to obtain 1D coordinates along the shower/track direction.
    \item \textbf{Find start}: Identify the minimum projection value, corresponding to the start of the cluster.
    \item \textbf{Extrapolate}: The vertex is the point on the axis at the minimum projection:
    \begin{equation}
        \vec{v}_\text{reco} = \bar{\vec{h}} + t_\text{min} \cdot \hat{e}_1, \quad t_\text{min} = \min_i \left[(\vec{h}_i - \bar{\vec{h}}) \cdot \hat{e}_1\right]
    \end{equation}
    where $\bar{\vec{h}}$ is the hit centroid and $\hat{e}_1$ is the principal axis.
\end{enumerate}

This algorithm leverages the directional structure of showers/tracks, achieving sub-centimeter vertex resolution with 1~cm pitch for clusters with $N \gtrsim 50$ hits.

\textbf{Angular reconstruction}: Track/shower directions are reconstructed using the same PCA. The direction is taken as the eigenvector $\hat{e}_1$ corresponding to the largest eigenvalue of the hit position covariance matrix.

%==============================================================================
\section{Ghost Hit Suppression}
\label{sec:ghosts}
%==============================================================================

\subsection{Analytical Demonstration}

We derive analytical expressions for ghost hit formation from first principles, building from the fundamental geometry of fiber readout. These derivations clarify when ghost contamination is most severe and quantify the advantage of the 3-View geometry.

\subsubsection{Definitions}

\textbf{Voxelized detector:} Consider a cubic detector volume divided into voxels of pitch $p$. Each voxel is indexed by integer coordinates $(i, j, k)$ corresponding to physical position:
\begin{equation}
    (x, y, z) = (i \cdot p, \, j \cdot p, \, k \cdot p)
\end{equation}

\textbf{Fiber geometry:} Three sets of orthogonal fibers read out the detector:
The detector is read out by three sets of orthogonal fibers. X-fibers run parallel to the X-axis, labeled by their $(y, z)$ or $(j, k)$ position. Y-fibers run parallel to the Y-axis, labeled by $(x, z)$ or $(i, k)$. Finally, Z-fibers run parallel to the Z-axis, labeled by $(x, y)$ or $(i, j)$.

\textbf{Hit registration:} When a particle traverses voxel $(i, j, k)$, three fibers register light:
When a particle traverses voxel $(i, j, k)$, three fibers register light: the X-fiber at $(j, k)$, the Y-fiber at $(i, k)$, and the Z-fiber at $(i, j)$.

\subsubsection{Ghost Formation in 2-View Geometry}

\begin{figure}[H]
    \centering
    \includegraphics[width=0.8\linewidth]{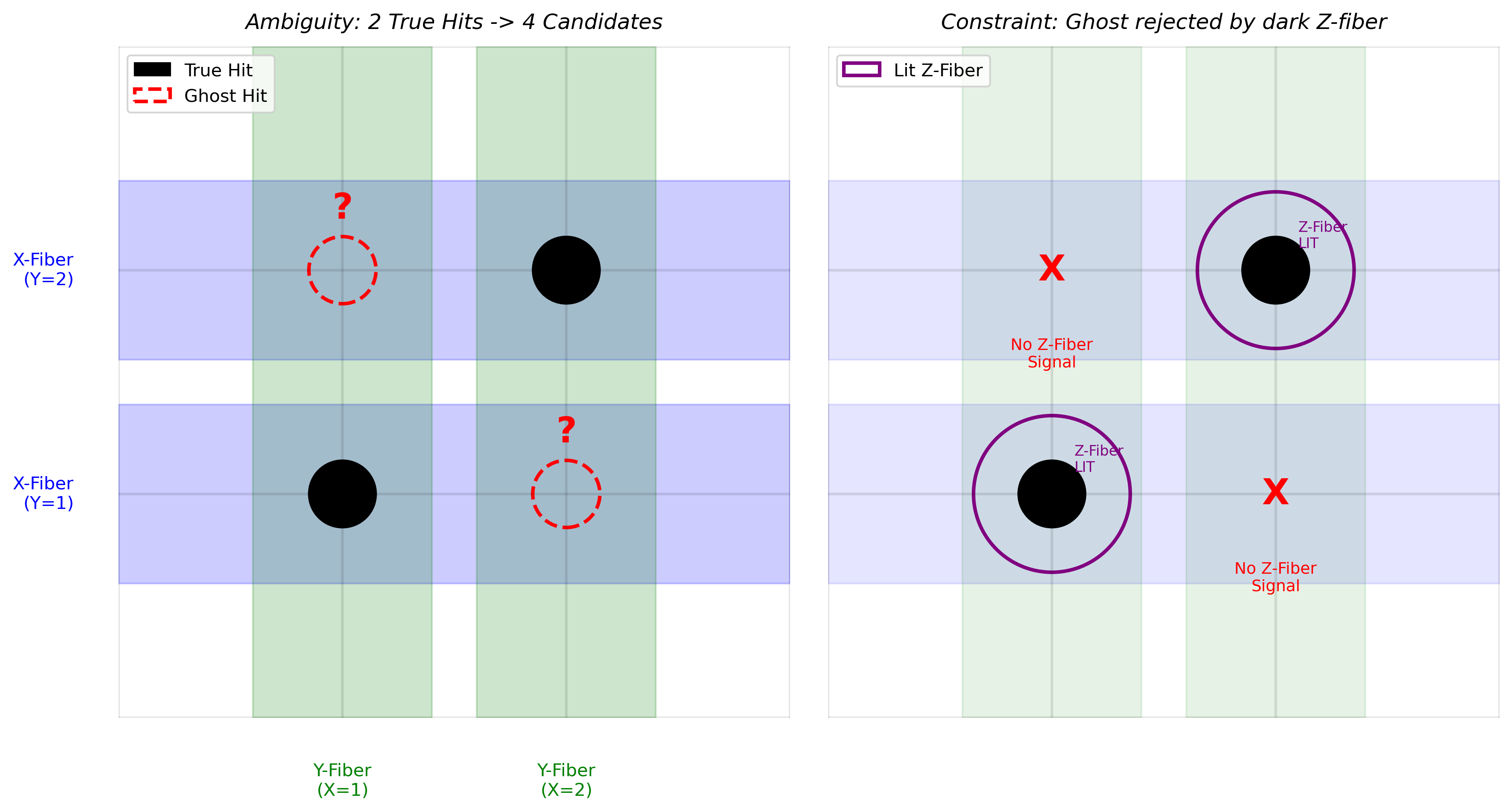}
    \caption{Schematic of ghost hit formation. Left: 2-View case showing ambiguous intersections. Right: 3-View case with Z-fiber veto eliminating ghosts.}
    \label{fig:ghost_mechanism}
\end{figure}

In a 2-View detector, only X and Y fibers are instrumented. The reconstruction algorithm identifies candidate hits by finding coincidences between lit X and Y fibers at the same $z$-layer. Figure~\ref{fig:ghost_mechanism} illustrates the ghost hit formation mechanism schematically. In the figure, although only two true hits are present, four total hits are reconstructed.

\medskip
\textbf{Reconstruction rule (2V)}

At $z$-layer $k$, let $\mathcal{X}_k = \{j_1, j_2, \ldots\}$ be the set of lit X-fiber indices, and $\mathcal{Y}_k = \{i_1, i_2, \ldots\}$ be the set of lit Y-fiber indices. The 2V candidate set is their Cartesian product:
\begin{equation}
    C_k^\text{2V} = \mathcal{Y}_k \times \mathcal{X}_k = \{(i, j) : i \in \mathcal{Y}_k, \, j \in \mathcal{X}_k\}
\end{equation}

\medskip
\textbf{Example}

Suppose two true hits exist at $(1, 3, 5)$ and $(2, 4, 5)$ (same $z$-layer $k=5$).
The lit X-fibers are $\mathcal{X}_5 = \{3, 4\}$ (corresponding to $y = 3p$ and $y = 4p$), while the lit Y-fibers are $\mathcal{Y}_5 = \{1, 2\}$ (corresponding to $x = p$ and $x = 2p$). This results in the candidate set $C_5^\text{2V} = \{(1,3), (1,4), (2,3), (2,4)\}$.

The candidates $(1,3)$ and $(2,4)$ correspond to true hits. The candidates $(1,4)$ and $(2,3)$ are \textbf{ghost hits}, i.e., positions where no energy was deposited, but the fiber coincidence creates an apparent hit.

\medskip
\textbf{Ghost count formula}

If $n_x = |\mathcal{X}_k|$ X-fibers and $n_y = |\mathcal{Y}_k|$ Y-fibers are lit in layer $k$, the number of candidates is $n_x \times n_y$. If all lit fibers arose from $N_\text{true}$ distinct true hits (where $N_\text{true} = n_x = n_y$ in the case of no overlapping projections), then:
\begin{equation}
    N_\text{candidates}^\text{2V} = n_x \times n_y, \quad N_\text{ghosts}^\text{2V} = n_x \times n_y - N_\text{true}
\end{equation}

For $N_\text{true}$ hits with distinct $(x, y)$ positions in a single layer:
\begin{equation}
    N_\text{ghosts}^\text{2V} = N_\text{true}^2 - N_\text{true} = N_\text{true}(N_\text{true} - 1)
    \label{eq:ghost_2v}
\end{equation}

\subsubsection{Ghost Suppression in 3-View Geometry}

In a 3-View detector, Z-fibers provide an additional constraint.

\medskip
\textbf{Reconstruction rule (3V)}

A candidate at $(i, j, k)$ is accepted only if the X-fiber $(j, k)$, the Y-fiber $(i, k)$, and the Z-fiber $(i, j)$ are all lit.

\medskip
\textbf{Ghost rejection mechanism}

A ghost at position $(i', j', k)$ generated by true hits at $(i_1, j_1, k)$ and $(i_2, j_2, k)$ (where $i' = i_1$, $j' = j_2$) would require Z-fiber $(i_1, j_2)$ to be lit. This Z-fiber is lit only if a true hit exists somewhere along the Z-axis at $(x, y) = (i_1 p, j_2 p)$. For a generic event, this is unlikely.

\medskip
\textbf{3V ghost condition}

A ghost survives at $(i', j', k)$ if and only if there exists some $k''$ such that $(i', j', k'')$ is a true hit (lighting the required Z-fiber).

\medskip
\textbf{Irreducible ghost configuration}

Consider four true hits forming a tetrahedron within a 2×2×2 box of voxels:
\begin{align}
    &(i_1, j_1, k_1), \quad (i_1, j_2, k_2), \quad (i_2, j_1, k_2), \quad (i_2, j_2, k_1)
\end{align}

\begin{figure}[H]
    \centering
    \includegraphics[width=0.8\linewidth]{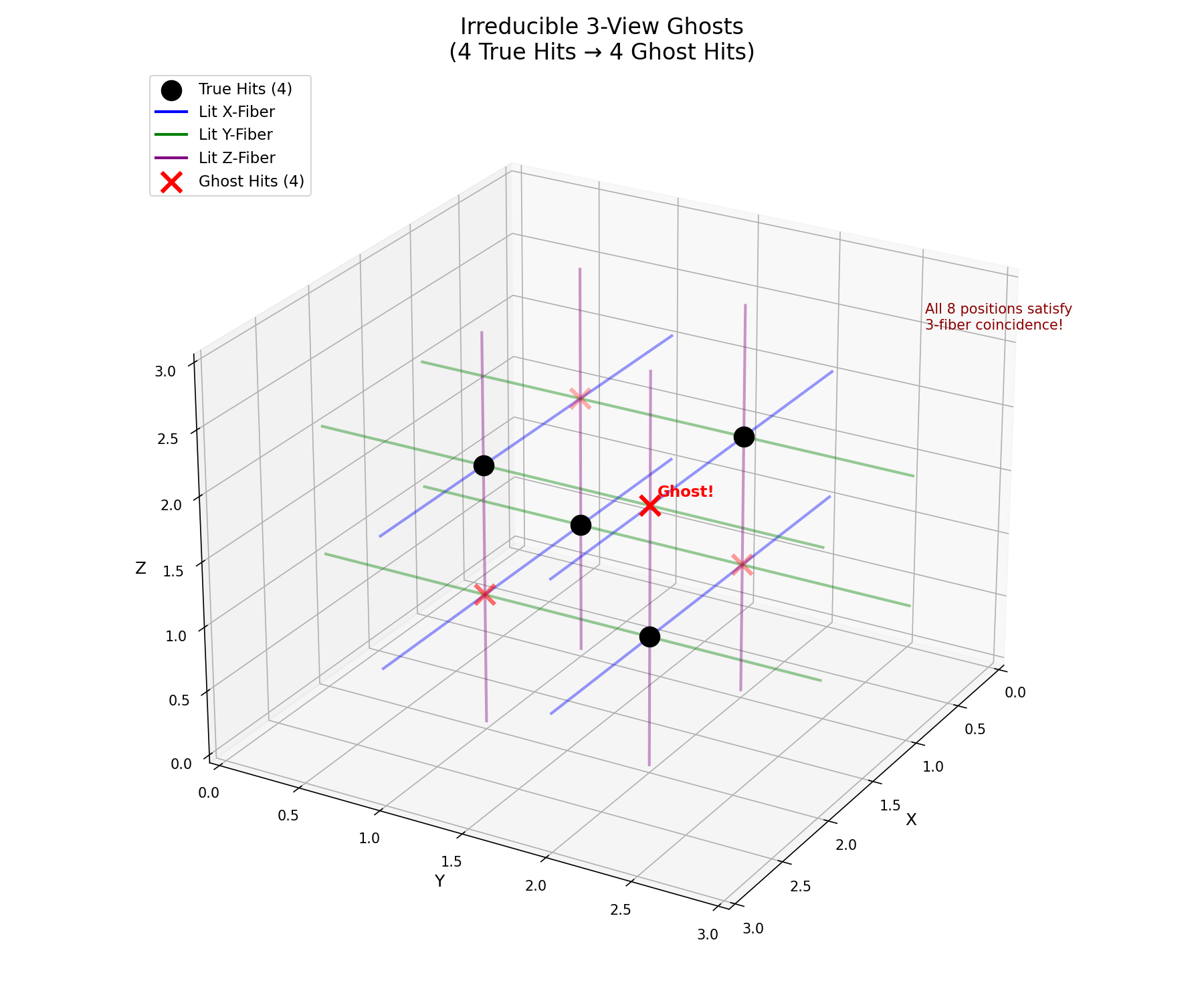}
    \caption{Example of irreducible ghosts in 3-View geometry. A configuration of four true hits creates fiber coincidences that satisfy the 3-View constraint at additional ghost positions.}
    \label{fig:ghost_3d}
\end{figure}

Here we use only two values for each index: $(i_1, i_2)$, $(j_1, j_2)$, $(k_1, k_2)$. The Z-fibers lit are all four combinations: $(i_1, j_1)$, $(i_1, j_2)$, $(i_2, j_1)$, $(i_2, j_2)$.
Now consider the position $(i_2, j_1, k_1)$, which is NOT a true hit. However, the X-fiber $(j_1, k_1)$ is lit by hit 1 at $(i_1, j_1, k_1)$, the Y-fiber $(i_2, k_1)$ is lit by hit 4 at $(i_2, j_2, k_1)$, and the Z-fiber $(i_2, j_1)$ is lit by hit 3 at $(i_2, j_1, k_2)$.
Therefore, this ghost at $(i_2, j_1, k_1)$ passes the 3V test. Similarly, there are four ghost positions in this configuration, corresponding to the remaining four corners of the 2×2×2 cube.
This configuration, having four true hits at alternating corners of a cube, creates \textbf{irreducible ghosts} that 3V cannot eliminate.
Figure \ref{fig:ghost_3d} illustrates such an irreducible ghost topology.

\subsubsection{Case 1: Track-Like Topology (Line) with $(\theta, \phi)$ Dependence}

Consider a straight track of length $L$ passing through $N$ voxels. The track direction is characterized by the polar angle $\theta$ from the Z-axis ($0^\circ$ corresponding to the Z-direction, $90^\circ$ to the XY plane) and the azimuthal angle $\phi$ in the XY plane ($0^\circ$ along X, $90^\circ$ along Y).

The direction vector is:
\begin{equation}
    \hat{d} = (\sin\theta \cos\phi, \; \sin\theta \sin\phi, \; \cos\theta),
\end{equation}
in which one should notice that the \textbf{ghost hit formation depends on BOTH $\theta$ AND $\phi$.}

\medskip
\textbf{Case A: Track along a principal axis ($\phi = 0^\circ$ or $90^\circ$, any $\theta$)}

When $\phi = 0^\circ$, the track lies in the XZ plane with direction $\hat{d} = (\sin\theta, 0, \cos\theta)$. All hits share the same $y$-coordinate (same Y-fiber for each $z$-layer). Since each $z$-layer has only one $(x, y)$ pair, the Cartesian product produces no ghosts:
\begin{equation}
    N_\text{ghost}^\text{2V}(\phi = 0^\circ \text{ or } 90^\circ) = 0 \quad \text{(for any $\theta$)}.
\end{equation}

This explains the paradoxical result that even a horizontal track ($\theta = 90^\circ$) produces no ghosts \emph{if} it travels along a fiber axis.

\medskip
\textbf{Case B: Track at diagonal azimuth ($\phi = 45^\circ$, $\theta = 90^\circ$)}

This is the \textbf{worst case} for ghost formation. The track lies in the XY plane at 45° to both axes:
\begin{equation}
    \hat{d} = \left(\frac{1}{\sqrt{2}}, \frac{1}{\sqrt{2}}, 0\right)
\end{equation}

A track of length $L$ crossing $N$ voxels at this orientation has all hits at the same $z$-coordinate. The $x$ and $y$ coordinates both vary along the track, creating $N$ distinct X-fibers and $N$ distinct Y-fibers. The 2V candidate reconstruction forms the complete Cartesian product:
\begin{equation}
    N_\text{candidates}^\text{2V} = N \times N = N^2
\end{equation}
\begin{equation}
    N_\text{ghost}^\text{2V}(\theta = 90^\circ, \phi = 45^\circ) = N^2 - N = N(N-1)
\end{equation}

For $N = 20$ hits, this produces $20 \times 19 = 380$ ghost hits, showing a dense ``wall'' of false candidates.

\medskip
\textbf{General formula for $\theta = 90^\circ$}

For a horizontal track at azimuthal angle $\phi$, the number of distinct X-fibers is proportional to the $y$-extent, and the number of distinct Y-fibers is proportional to the $x$-extent:
\begin{align}
    n_X &= \text{(hits with distinct $y$)} \propto L |\sin\phi| / p,  \\
    n_Y &= \text{(hits with distinct $x$)} \propto L |\cos\phi| / p.
\end{align}

The ghost count scales as:
\begin{equation}
\begin{split}
    N_\text{ghost}^\text{2V}(\theta = 90^\circ, \phi) &\approx n_X \cdot n_Y - N \\
    &\propto \frac{L^2 |\sin\phi \cos\phi|}{p^2} = \frac{L^2 |\sin 2\phi|}{2p^2}.
\end{split}
\end{equation}

This function is maximized at $\phi = 45^\circ$ (where $|\sin 2\phi| = 1$) and zero at $\phi = 0^\circ$ or $90^\circ$.

\medskip
\textbf{General formula for arbitrary $(\theta, \phi)$}

For a track oriented at general angles $(\theta, \phi)$, we derive the ghost count by analyzing the fiber overlaps in each $z$-layer. The track direction is $\hat{d} = (\sin\theta\cos\phi, \sin\theta\sin\phi, \cos\theta)$.

Consider a track segment of length $L$ starting at the origin. The endpoint is at:
\begin{equation}
    (x_\text{end}, y_\text{end}, z_\text{end}) = L(\sin\theta\cos\phi, \sin\theta\sin\phi, \cos\theta).
\end{equation}

The track spans:
\begin{align}
    \Delta x &= L \sin\theta |\cos\phi|, \\
    \Delta y &= L \sin\theta |\sin\phi|, \\
    \Delta z &= L \cos\theta.
\end{align}

\medskip
\textbf{Z-layer analysis}

The number of $z$-layers spanned is $N_z = \Delta z / p = L\cos\theta / p$. Within each $z$-layer of thickness $p$, the track traverses:
\begin{align}
    \delta x &= \frac{\Delta x}{N_z} = p \tan\theta |\cos\phi|, \\
    \delta y &= \frac{\Delta y}{N_z} = p \tan\theta |\sin\phi|.
\end{align}

The number of voxels crossed per $z$-layer is approximately,
\begin{equation}
    n_\text{vox/layer} \approx \max\left(1, \frac{\delta x}{p}\right) \times \max\left(1, \frac{\delta y}{p}\right) = \max(1, \tan\theta|\cos\phi|) \times \max(1, \tan\theta|\sin\phi|).
\end{equation}

\medskip
\textbf{Ghost count per layer}

For $n$ hits in a single $z$-layer with $n_x$ distinct $x$-values and $n_y$ distinct $y$-values:
\begin{equation}
    N_\text{ghost/layer} = n_x \cdot n_y - n.
\end{equation}

For a straight track, $n_x \approx \delta x / p$ and $n_y \approx \delta y / p$, so:
\begin{equation}
    N_\text{ghost/layer} \approx \tan^2\theta \cdot |\sin\phi\cos\phi| - 1 = \frac{\tan^2\theta \cdot |\sin 2\phi|}{2} - 1.
\end{equation}

\medskip
\textbf{Total 2V ghost count}

Summing over all $N_z$ layers:
\begin{equation}
    \begin{aligned}
    N_\text{ghost}^\text{2V}(\theta, \phi) &\approx \frac{L\cos\theta}{p} \cdot \left(\frac{\tan^2\theta \cdot |\sin 2\phi|}{2}\right) \\
    &= \frac{L \sin^2\theta \cdot |\sin 2\phi|}{2p\cos\theta}.
    \end{aligned}
    \label{eq:ghost_general}
\end{equation}

This can be rewritten as:
\begin{equation}
    N_\text{ghost}^\text{2V}(\theta, \phi) \propto \frac{\sin^2\theta}{\cos\theta} \cdot |\sin 2\phi| = \tan\theta \sin\theta \cdot |\sin 2\phi|.
\end{equation}

\medskip
\textbf{Limiting cases}
\begin{itemize}
    \item $\theta = 0^\circ$: $\sin\theta = 0 \Rightarrow N_\text{ghost} = 0$ (track along Z, no XY spread);
    \item $\phi = 0^\circ$ or $90^\circ$: $|\sin 2\phi| = 0 \Rightarrow N_\text{ghost} = 0$ (track along fiber axis);
    \item $\theta \to 90^\circ$, $\phi = 45^\circ$: Maximum ghosts, formula diverges as track becomes horizontal diagonal.
\end{itemize}

\medskip
\textbf{3-View ghost count (track)}

For a straight track, the Z-fiber at $(x_i, y_j)$ is lit only if a true hit exists at that $(x, y)$ position. Since each hit has a unique $(x, y)$, ghosts at $(x_i, y_j)$ for $i \neq j$ have no corresponding lit Z-fiber:
\begin{equation}
    N_\text{ghost}^\text{3V,track} = 0 \quad \text{(for any single straight track, any $\theta$, any $\phi$)}.
\end{equation}

In summary by far, we see that the 3-View geometry provides complete ghost suppression for single straight tracks at all orientations, while 2-View ghost contamination follows $N_\text{ghost} \propto \tan\theta\sin\theta \cdot |\sin 2\phi|$, peaking at $(\theta, \phi) = (90^\circ, 45^\circ)$.

\subsubsection{Case 2: Shower-Like Topology (Cone)}

Consider a cone-shaped shower with apex at the origin (vertex), axis along the positive Z-direction ($\theta = 0^\circ$ initially), opening half-angle $\alpha$, length $L$, and $N$ hits uniformly distributed within the cone volume.

\medskip
\textbf{Step 1: Cone geometry}

At distance $t$ from the apex along the axis, the cone radius is:
\begin{equation}
    r(t) = t \tan\alpha.
\end{equation}

The volume element of a thin cylindrical slice at distance $t$ with thickness $dt$ is:
\begin{equation}
    dV = \pi r^2(t) \, dt = \pi t^2 \tan^2\alpha \, dt.
\end{equation}

\medskip
\textbf{Step 2: Total cone volume}
Integrating from apex ($t=0$) to end ($t=L$):
\begin{equation}
    V_\text{cone} = \int_0^L \pi t^2 \tan^2\alpha \, dt = \pi \tan^2\alpha \cdot \frac{L^3}{3} = \frac{\pi L^3 \tan^2\alpha}{3}.
\end{equation}

\medskip
\textbf{Step 3: Hit density}
With $N$ hits uniformly distributed in the volume, the hit density (hits per unit volume) is:
\begin{equation}
    \rho_\text{vol} = \frac{N}{V_\text{cone}} = \frac{3N}{\pi L^3 \tan^2\alpha}.
\end{equation}

The number of hits in a slice at position $t$ of thickness $dt$ is:
\begin{equation}
    dN = \rho_\text{vol} \cdot dV = \frac{3N}{\pi L^3 \tan^2\alpha} \cdot \pi t^2 \tan^2\alpha \, dt = \frac{3N t^2}{L^3} \, dt.
\end{equation}

Therefore, the \textbf{linear hit density} (hits per unit length along the axis) is:
\begin{equation}
    \rho(t) = \frac{dN}{dt} = \frac{3N t^2}{L^3}.
\end{equation}

\medskip
\textbf{Step 4: Hits per $z$-layer and ghost scaling}
For a cone along the Z-axis ($\theta = 0^\circ$), at distance $z$ from the apex, the cone has radius:
\begin{equation}
    r(z) = z \tan\alpha.
\end{equation}

The number of hits in a $z$-layer of thickness $p$ centered at $z$ is:
\begin{equation}
    n_z(z) = \rho(z) \cdot p = \frac{3Np \, z^2}{L^3}.
\end{equation}

\medskip
\textbf{Averaging over track orientations}

Consider these $n$ hits as forming radial lines at all azimuthal angles $\phi$. A track-like segment at angle $\phi$ contributes ghosts proportional to $|\sin\phi \cos\phi| = |\sin 2\phi|/2$. Averaging over all orientations:
\begin{equation}
    \langle |\sin\phi \cos\phi| \rangle = \frac{1}{2\pi} \int_0^{2\pi} |\sin\phi \cos\phi| \, d\phi = \frac{1}{\pi}
\end{equation}

Therefore, for $n$ hits uniformly distributed around a circle, the average ghost count per layer scales as:
\begin{equation}
    N_\text{ghost/layer} \propto \frac{n^2}{\pi} \quad \text{(hit-limited regime: } n \ll 2r/p \text{)}.
\end{equation}

However, when hits densely fill the circle ($n \gg 2r/p$), the fiber coverage saturates at the diameter:
\begin{equation}
    N_\text{ghost/layer} \propto \left(\frac{2r}{p}\right)^2 \quad \text{(fiber-limited regime)}.
\end{equation}

The unified scaling results in:
\begin{equation}
    N_\text{ghost/layer}(n, r) \propto \min\left(n^2, \left(\frac{r}{p}\right)^2\right).
\end{equation}

\medskip
\textbf{Total ghost count (integrating over cone layers)}

For a full cone with $r(z) = z\tan\alpha$ and layer hit count $n_z \propto z^2$:
\begin{equation}
    N_\text{ghost}^\text{2V,cone} \propto \int_0^L \min\left(n_z^2, \left(\frac{z \tan\alpha}{p}\right)^2\right) dz.
\end{equation}

The transition from hit-limited to fiber-limited scaling occurs when $n_z \sim r(z)/p$, i.e., when the hit density per layer exceeds the fiber resolution.

\medskip
\textbf{2-View ghost count (cone, general $\theta$ and $\phi$)}

Unlike a straight track, a cone has finite transverse extent due to its opening angle $\alpha$. This means even at $\phi = 0^\circ$ (cone axis in XZ plane), the cone hits spread in both $x$ and $y$ directions, creating ghost opportunities.
For a cone with axis at angles $(\theta, \phi)$, the projection onto the XY plane has an elliptical footprint. The $x$-extent and $y$-extent of the cone depend on both the cone orientation and the opening angle:
\begin{align}
    \Delta x_\text{cone} &\approx L\sin\theta|\cos\phi| + 2L\tan\alpha \cdot f_x(\theta, \phi), \\
    \Delta y_\text{cone} &\approx L\sin\theta|\sin\phi| + 2L\tan\alpha \cdot f_y(\theta, \phi),
\end{align}
where $f_x$ and $f_y$ are geometric factors of order unity that depend on the cone orientation.

For cones, the opening angle $\alpha$ provides a ``baseline'' spread that creates ghosts even when the axis aligns with a fiber direction. The key difference from tracks is that ghosts scale with \emph{geometry} ($L$, $\alpha$, $p$) rather than hit count $N$.

\medskip
\textbf{Simplified scaling}

For practical purposes, the 2V ghost count for a cone scales as:
\begin{equation}
    N_\text{ghost}^\text{2V,cone}(\theta, \phi) \propto \frac{L^3 \tan^2\alpha}{p^3 \cos\theta}.
\end{equation}

This is independent of hit count $N$ for well-sampled cones (when $N$ is large enough to fill the fiber diameter). The $1/\cos\theta$ factor accounts for layer compression as the cone tilts.
At $\theta \to 90^\circ$, all hits project into $\sim 1$ layer. In this limit, the fiber coverage becomes maximal:
\begin{equation}
    N_\text{ghost}^\text{2V,cone}(\theta \to 90^\circ) \to \left(\frac{2L\tan\alpha}{p}\right)^2.
\end{equation}

\medskip
\textbf{3-View ghost count (cone)}

A ghost at $(x_i, y_j, z)$ requires a lit Z-fiber at $(x_i, y_j)$. For a circular distribution, the probability that a random $(x, y)$ position coincides with a true hit's Z-fiber scales as:
\begin{equation}
    P_\text{Z-fiber} \sim \frac{N \cdot p^2}{\pi r^2} = \frac{Np^2}{\pi z^2 \tan^2\alpha}
\end{equation}.

The 3V ghost count is the 2V ghost count reduced by this probability:
\begin{equation}
    N_\text{ghost}^\text{3V,cone} \sim N_\text{ghost}^\text{2V} \times P_\text{Z-fiber} \sim \frac{L^3 \tan^2\alpha}{p^3} \times \frac{Np^2}{\pi L^2 \tan^2\alpha} = \frac{NL}{p\pi}.
\end{equation}

The \textbf{ghost reduction factor} is:
\begin{equation}
    R = \frac{N_\text{ghost}^\text{2V}}{N_\text{ghost}^\text{3V}} \sim \frac{L^2 \tan^2\alpha}{Np^2}
\end{equation}

For typical parameters ($N = 100$, $L = 20$ cm, $p = 1$ cm, $\alpha = 30^\circ$, $\tan 30^\circ \approx 0.58$):
\begin{equation}
    R \sim \frac{400 \times 0.33}{100 \times 1} \approx 1.3.
\end{equation}

This modest reduction factor indicates that for compact cones, 3V provides moderate (factor of $\sim$few) ghost suppression. The advantage increases for wider cones (larger $\alpha$) or longer clusters (larger $L/p$).

\subsubsection{Summary of Scaling Laws}

The ghost hit count in 2-View geometry depends on both the polar angle $\theta$ and the azimuthal angle $\phi$:

\begin{table}[H]
\centering
\begin{tabular}{|l|c|c|c|}
\hline
\textbf{Topology} & \textbf{Orientation} & \textbf{2V Ghosts} & \textbf{3V Ghosts} \\
\hline
Track & $\theta = 0^\circ$ (along Z) & $0$ & $0$ \\
Track & $\theta = 90^\circ$, $\phi = 0^\circ$ (along X) & $0$ & $0$ \\
Track & $\theta = 90^\circ$, $\phi = 45^\circ$ (diagonal) & $N(N-1)$ & $0$ \\
\hline
Cone & $\theta = 0^\circ$, any $\phi$ & $\propto L^3 \tan^2\alpha / p^3$ & $\propto NL/p$ \\
Cone & $\theta = 90^\circ$ & $(2L\tan\alpha/p)^2$ & Reduced by $R$ \\
\hline
\end{tabular}
\caption{Ghost hit counts depend on both topology and orientation. For cones, ghosts scale with geometry ($L$, $\alpha$, $p$) rather than hit count $N$. The 3-View reduction factor scales as $R \sim L^2\tan^2\alpha / (Np^2)$, where $L$ is the shower length, $\alpha$ is the cone half-opening angle, and $p$ is the voxel pitch.}
\label{tab:ghost_scaling}
\end{table}

In summary, for tracks, ghost contamination follows a $|\sin 2\phi|$ dependence, peaking at $\phi = 45^\circ$ and vanishing at principal axis orientations. For cones, ghost hits are determined by the fiber diameter coverage ($2r/p$), making them independent of hit count for well-sampled distributions.

\subsection{Quantitative Comparison}

\begin{figure}[H]
    \centering
    \includegraphics[width=1.0\linewidth]{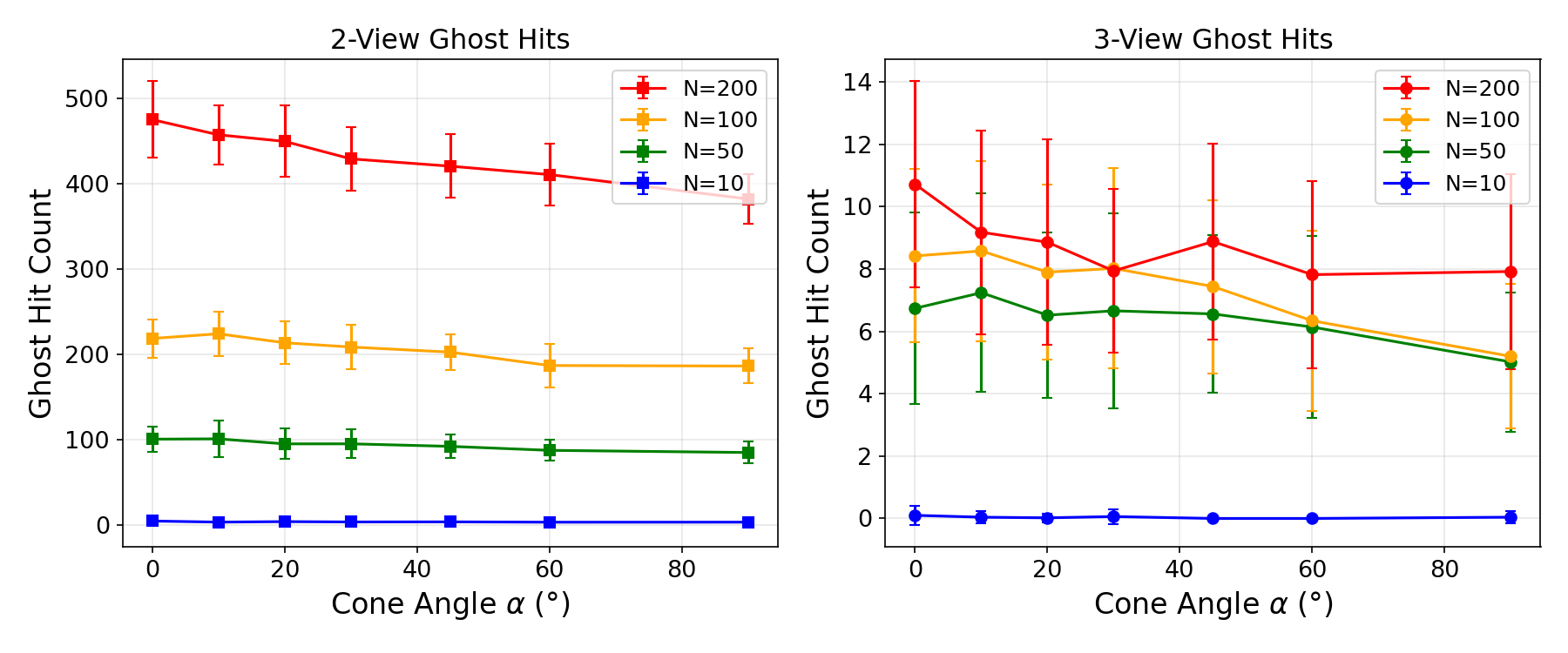}
    \caption{Ghost hit count vs cone angle $\alpha$ for $N = 10, 50, 100, 200$ hits. Left: 2-View. Right: 3-View. Shower parameters: $L = 20$~cm, orientations averaged over $\theta \in [0^\circ, 90^\circ]$ and $\phi \in [0^\circ, 90^\circ]$. Error bars show standard deviation across 50 trials.}
    \label{fig:ghost_study}
\end{figure}

Figure \ref{fig:ghost_study} shows the ghost hit count as a function of cone angle $\alpha$ for various cluster sizes $N$. The simulation uses $L = 20$ cm shower length, uniform orientation averaging over $\theta \in [0^\circ, 90^\circ]$ and $\phi \in [0^\circ, 90^\circ]$. The 3-View geometry achieves 50 to 95\% reduction in ghost hits compared to 2-View, with the largest gains for high-$N$ clusters.

\subsection{Orientation Dependence $(\theta, \phi)$ }

\begin{figure}[H]
    \centering
    \includegraphics[width=1.0\linewidth]{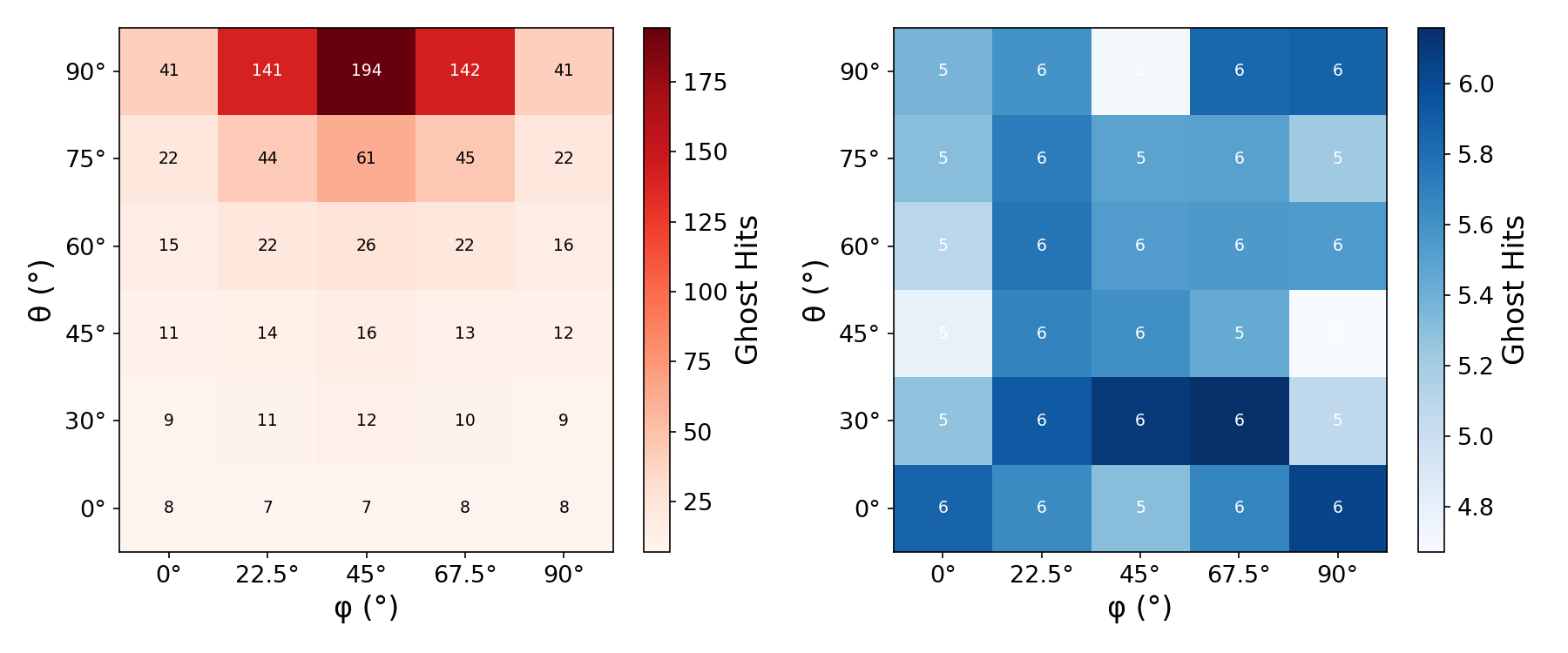}
    \caption{Ghost hit count heatmap for $\alpha = 5^\circ$ (track-like). Left: 2-View. Right: 3-View. Note the strong $\phi$ dependence at large $\theta$, matching the $|\sin 2\phi|$ analytical prediction.}
    \label{fig:ghost_heatmap_5}
\end{figure}

\begin{figure}[H]
    \centering
    \includegraphics[width=1.0\linewidth]{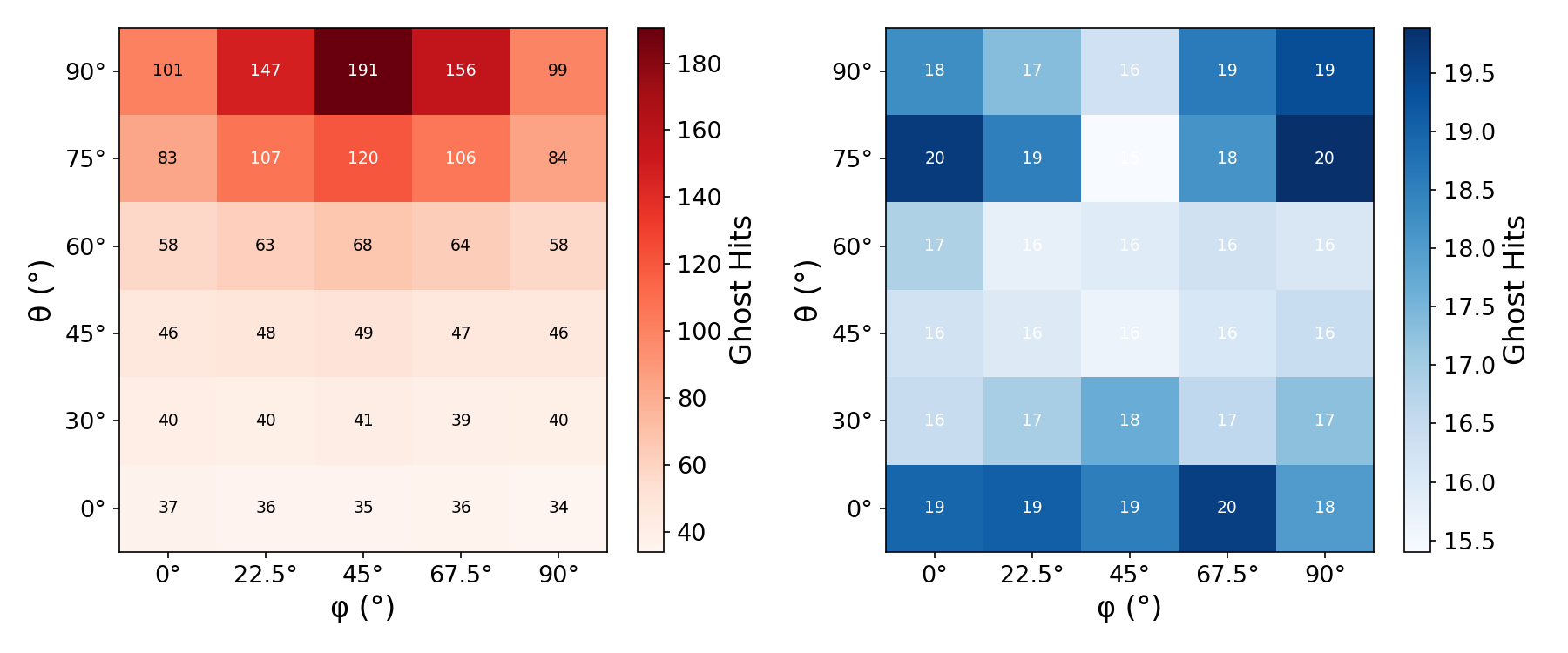}
    \caption{Ghost hit count heatmap for $\alpha = 15^\circ$ (narrow cone). Left: 2-View. Right: 3-View.}
    \label{fig:ghost_heatmap_15}
\end{figure}

\begin{figure}[H]
    \centering
    \includegraphics[width=1.0\linewidth]{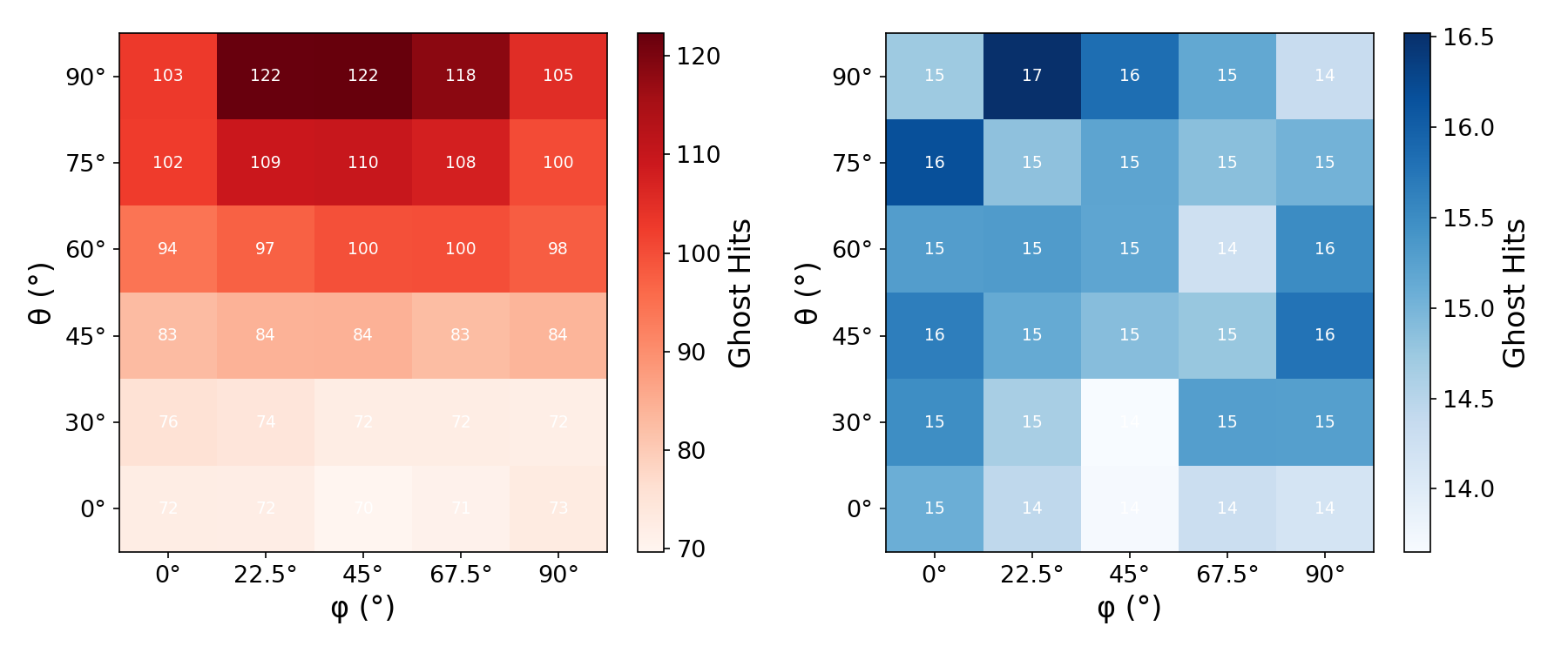}
    \caption{Ghost hit count heatmap for $\alpha = 40^\circ$ (wide cone). Left: 2-View. Right: 3-View. The $\phi$ dependence is weaker due to the cone's intrinsic transverse spread.}
    \label{fig:ghost_heatmap_40}
\end{figure}

Figures~\ref{fig:ghost_heatmap_5} to \ref{fig:ghost_heatmap_40} show the ghost hit count as a function of both polar angle $\theta$ and azimuthal angle $\phi$ for various cone opening angles. In each figure, the left panel shows the 2-View result and the right panel shows the 3-View result. These heatmaps confirm the analytical prediction that 2V ghost contamination follows the $|\sin 2\phi|$ pattern, with maximum ghosts at $\phi = 45^\circ$ and near-zero at $\phi = 0^\circ$ or $90^\circ$ (fiber axis orientations).

\begin{figure}[H]
    \centering
    \includegraphics[width=1.0\linewidth]{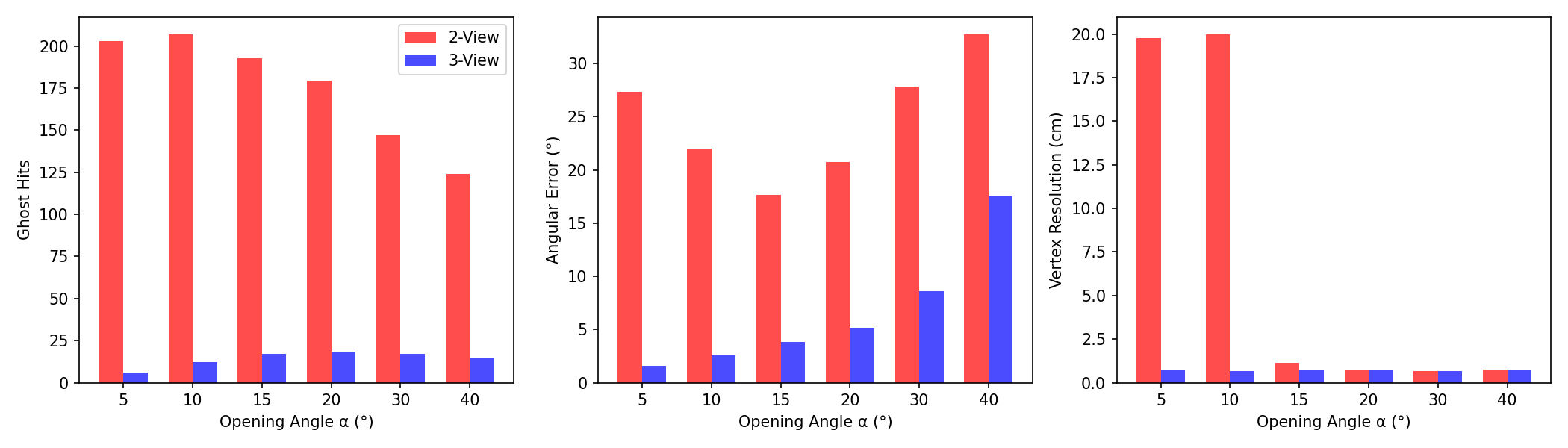}
    \caption{Summary comparison at worst-case orientation $(\theta = 90^\circ, \phi = 45^\circ)$. Parameters: $N = 50$ hits, $L = 20$~cm, opening angles $\alpha = 5^\circ, 10^\circ, 15^\circ, 20^\circ, 30^\circ, 40^\circ$, 100 trials per configuration. Bar charts show ghost count (left), angular resolution (middle), and vertex resolution (right) for 2V (red) and 3V (blue) across different opening angles. The 3V advantage is dramatic, especially for narrow cones.}
    \label{fig:summary_worst}
\end{figure}

Figure~\ref{fig:summary_worst} summarizes the performance comparison at the worst-case orientation $(\theta = 90^\circ, \phi = 45^\circ)$ for different cone opening angles.

%==============================================================================
\section{Single-Shower Angular Resolution}
\label{sec:angular}
%==============================================================================

Angular resolution for electromagnetic showers is critical for $\pi^0$ reconstruction. In neutral-current $\pi^0$ production events, the $\pi^0$ decays to two photons whose directions and energies determine the $\pi^0$ kinematics. Mis-reconstruction of shower directions leads to systematic shifts in the reconstructed invariant mass of neutral pion candidates and degraded background rejection.

We define the \textit{shower linearity} $\lambda$ as the fraction of total variance captured by the first principal component of the hit distribution:
\begin{equation}
    \lambda = \frac{\sigma_1}{\sigma_1 + \sigma_2 + \sigma_3},
\end{equation}
where $\sigma_i$ are the eigenvalues of the hit position covariance matrix, sorted in descending order. A perfectly collinear track has $\lambda = 1$, while a spatially isotropic distribution has $\lambda \approx 1/3$.

We note that hit amplitude (energy deposition per voxel) could provide an additional handle for ghost rejection, but is not used here since we focus on the geometric readout topology alone.

\begin{figure}[H]
    \centering
    \includegraphics[width=1.0\linewidth]{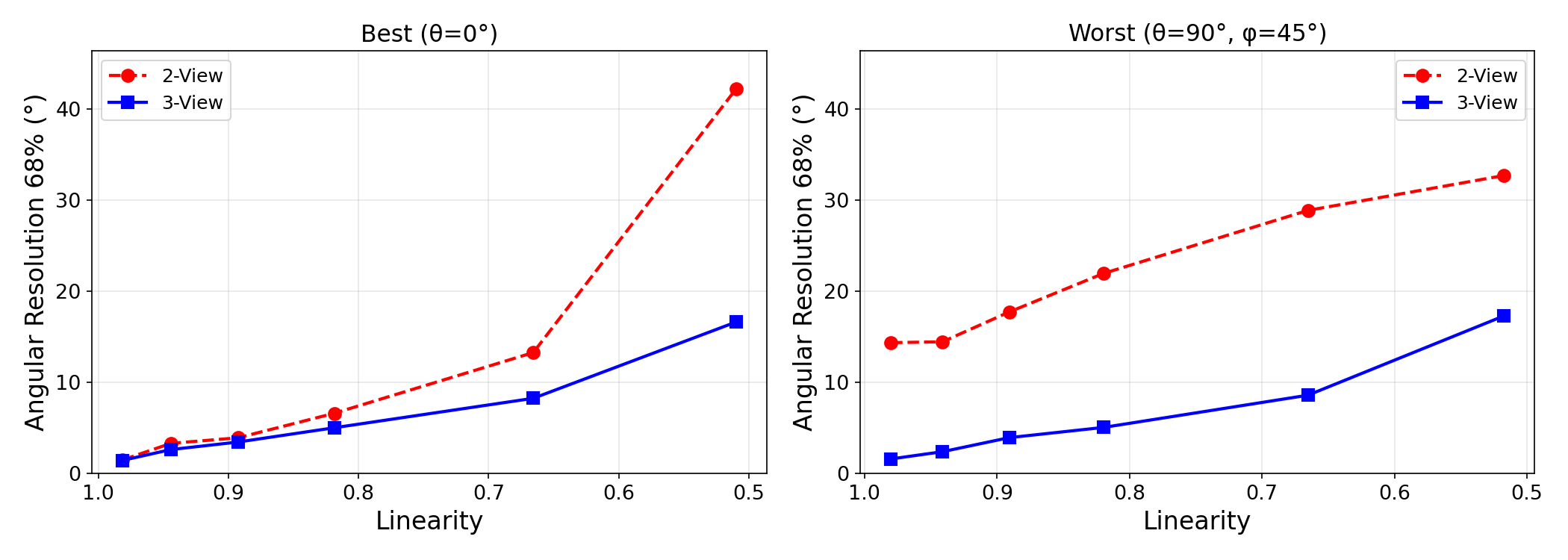}
    \caption{Single-shower angular resolution (68th percentile) vs linearity for $N=50$ hits, $L=20$ cm. Left: best case ($\theta = 0^\circ$, $\phi = 0^\circ$). Right: worst case ($\theta = 90^\circ$, $\phi = 45^\circ$). Blue (3V) consistently matches or beats red (2V), with the largest advantage at worst-case orientation.}
    \label{fig:angular_res}
\end{figure}

We generate cone-shaped hit patterns with $N = 20, 50, 100$ hits, lengths $L = 10, 20, 40$ cm, and opening angles from 10$^\circ$ to 75$^\circ$. Figure \ref{fig:angular_res} shows the 68th percentile angular resolution as a function of shower linearity. The 3-View geometry consistently matches or outperforms 2-View, with the largest improvements for high-linearity showers where ghost hits create a spurious ``halo'' that biases the PCA direction.

\begin{figure}[H]
    \centering
    \includegraphics[width=0.9\linewidth]{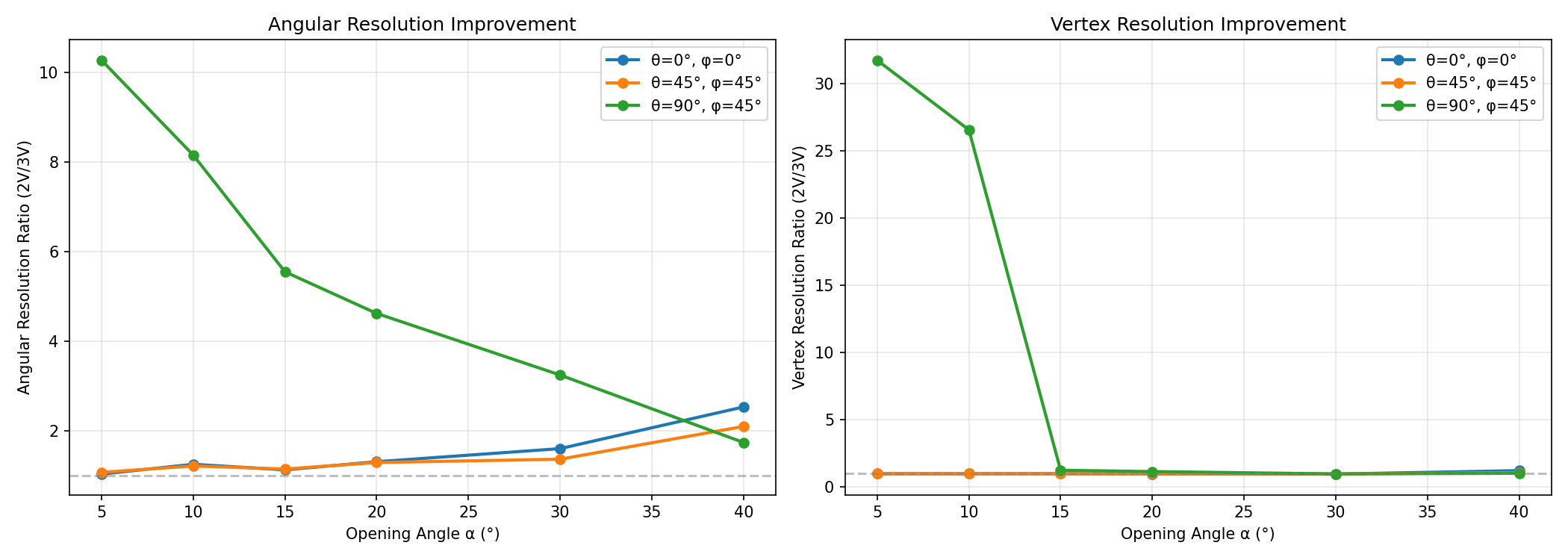}
    \caption{Improvement ratio (2V resolution / 3V resolution) for angular (left) and vertex (right) reconstruction. Parameters: $N=50$ hits, $L=20$ cm. Curves show three orientations: $\theta=0^\circ$ (best), $\theta=45^\circ$, and $\theta=90^\circ$ (worst). Ratios $> 1$ indicate 3V advantage.}
    \label{fig:angular_improve}
\end{figure}

Figure~\ref{fig:angular_improve} shows the improvement ratio (2V/3V resolution) for both angular and vertex reconstruction. At the worst-case orientation ($\theta = 90^\circ$, $\phi = 45^\circ$), the 3V advantage reaches 5 to 10$\times$ for track-like showers and decreases toward unity for isotropic distributions where ghosts are less problematic.

\subsection{Angular Resolution $(\theta, \phi)$ Dependence}

\begin{figure}[H]
    \centering
    \includegraphics[width=1.0\linewidth]{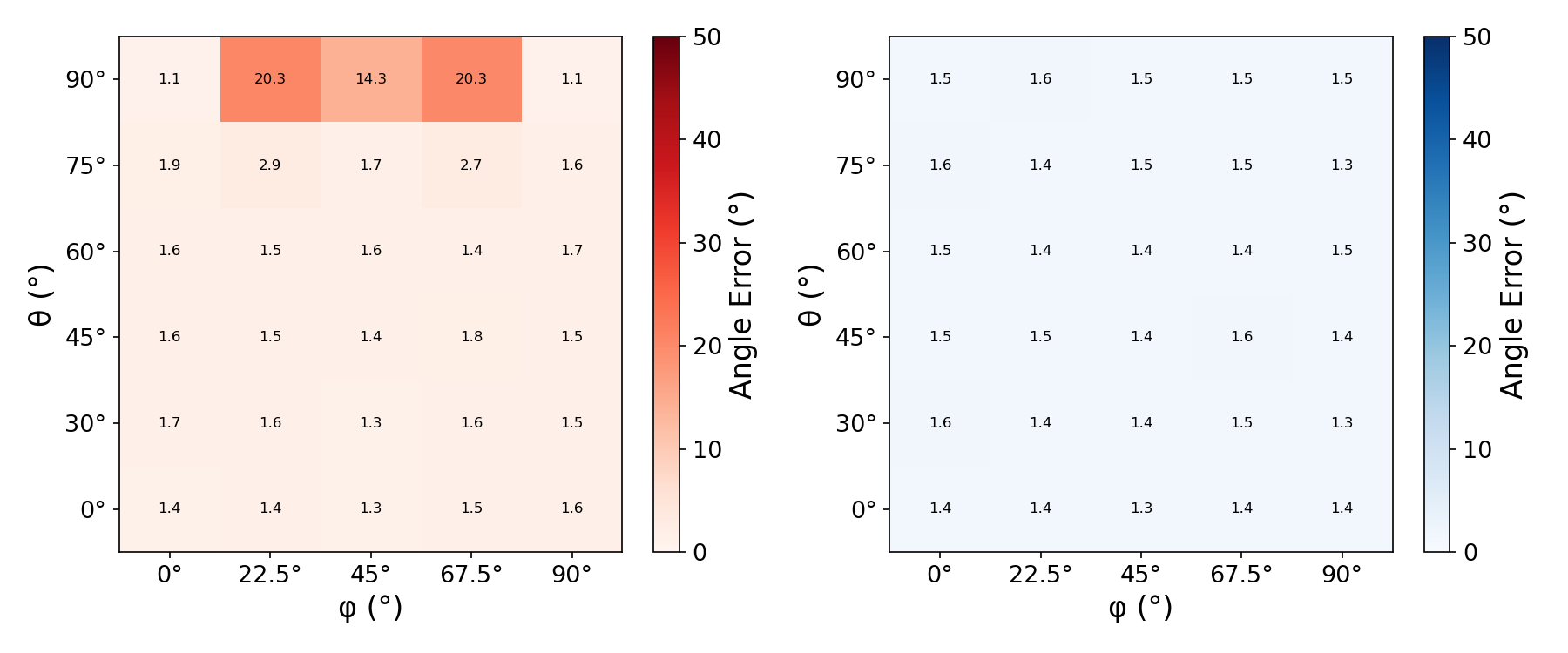}
    \caption{Angular resolution heatmap ($N=50$ hits, $L=20$ cm) for $\alpha = 5^\circ$ (track-like). Left: 2-View. Right: 3-View. The 2V resolution degrades dramatically at $(\theta = 90^\circ, \phi = 45^\circ)$ while 3V remains stable across all orientations.}
    \label{fig:angular_heatmap_5}
\end{figure}

\begin{figure}[H]
    \centering
    \includegraphics[width=1.0\linewidth]{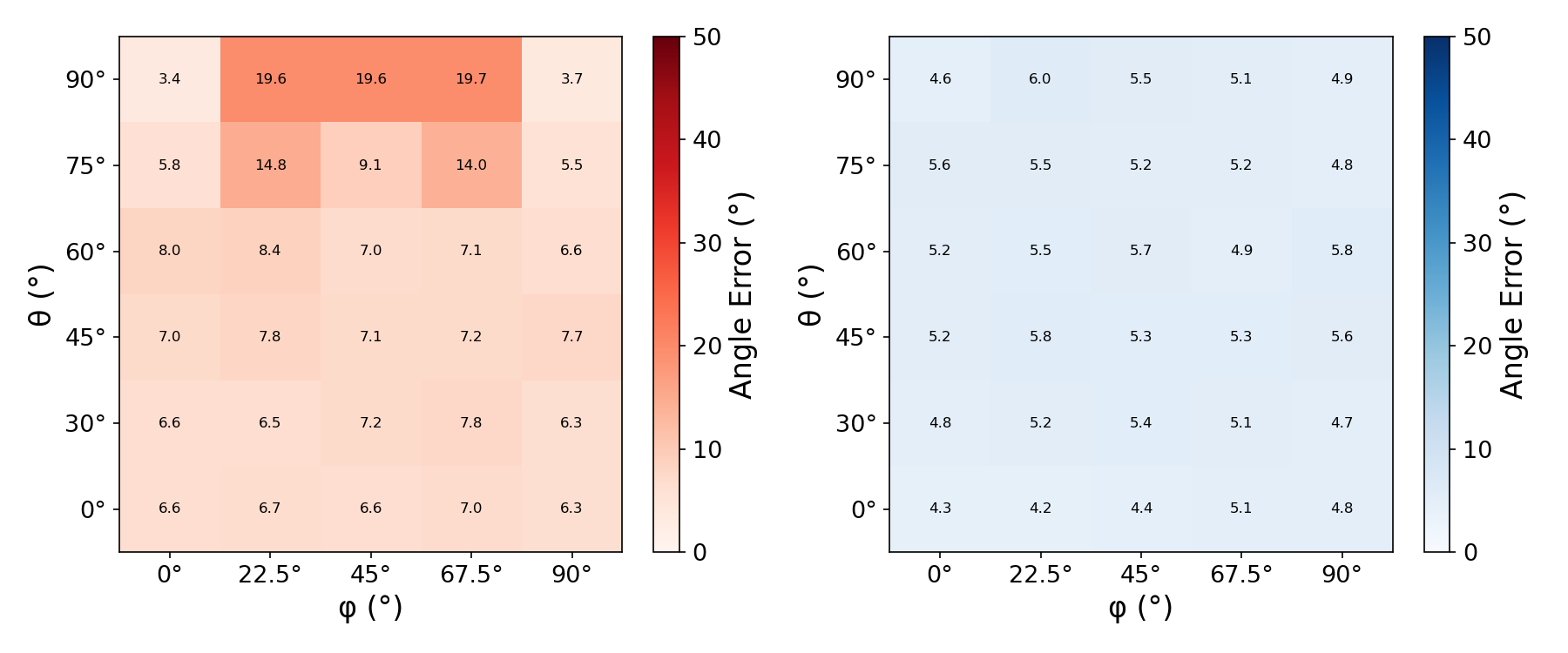}
    \caption{Angular resolution heatmap ($N=50$ hits, $L=20$ cm) for $\alpha = 20^\circ$ (moderate cone). Left: 2-View. Right: 3-View. Both 2V and 3V show some orientation dependence.}
    \label{fig:angular_heatmap_20}
\end{figure}

Figures~\ref{fig:angular_heatmap_5} to \ref{fig:angular_heatmap_20} show the angular resolution as a function of shower orientation for track-like ($\alpha = 5^\circ$) and shower-like ($\alpha = 20^\circ$) topologies. In each figure, the left panel shows the 2-View result and the right panel shows the 3-View result.

In Figure~\ref{fig:angular_heatmap_5} (2-View, left panel), the worst angular resolution occurs not at $\phi = 45^\circ$ but at nearby angles ($\phi \approx 22.5^\circ$ and $67.5^\circ$). This is because at exactly $\phi = 45^\circ$, the ghost hits form a symmetric distribution around the true track axis, which biases the PCA direction less than an asymmetric ghost distribution at neighboring $\phi$ values. At off-diagonal angles, the asymmetric ghost distribution pulls the PCA axis away from the true direction more effectively.

In Figure~\ref{fig:angular_heatmap_20} (3-View, right panel), the small variations in angular resolution across orientations are consistent with statistical fluctuations from the finite number of trials (100 per point). The 3-View geometry is effectively orientation-independent, with cell-to-cell variations of order 1$^\circ$, within statistical uncertainty.

%==============================================================================
\section{Single-Shower Vertex Resolution}
\label{sec:single_vertex}
%==============================================================================

Locating the start point of an electromagnetic shower is important for photon conversion point reconstruction. In events with photons converting in the detector material ($\gamma \to e^+ e^-$), the conversion distance from the interaction vertex helps distinguish prompt photons from $\pi^0$ decay products.

\subsection{Vertex Finding Algorithm}

The single-shower vertex finding algorithm exploits the geometric property that an electromagnetic shower or hadronic track originates from a localized vertex and expands outward. We employ a PCA approach with a novel \emph{transverse spread disambiguation} technique to identify the vertex end of an elongated hit cluster.

\paragraph{Step 1: PCA Axis Extraction}

Given a set of $N$ hit positions $\{\vec{r}_i\}$, we compute the centroid $\bar{\vec{r}} = \frac{1}{N}\sum_i \vec{r}_i$ and the covariance matrix:
\begin{equation}
    C_{jk} = \frac{1}{N-1} \sum_{i=1}^N (r_{ij} - \bar{r}_j)(r_{ik} - \bar{r}_k).
\end{equation}
The principal axis $\hat{a}$ is the eigenvector corresponding to the largest eigenvalue $\lambda_1$. For a shower or track, this axis represents the average direction of propagation.

\paragraph{Step 2: Projection onto Principal Axis}

Each hit is projected onto the principal axis:
\begin{equation}
    t_i = (\vec{r}_i - \bar{\vec{r}}) \cdot \hat{a}.
\end{equation}
The projection $t_i$ measures the signed distance along the shower axis from the centroid. The two extremal hits are those with minimum and maximum projections: $\vec{r}_{\min} = \vec{r}_{i : t_i = t_{\min}}$ and $\vec{r}_{\max} = \vec{r}_{i : t_i = t_{\max}}$.

\paragraph{Step 3: Transverse Spread Disambiguation}

The key challenge is determining which extremum corresponds to the vertex. For a cone-shaped shower, the vertex (apex) has \emph{low transverse spread}, such that the hits are concentrated near the axis, while the far end (base) has \emph{high transverse spread}, with hits spread outward in a disk.

We split the hits into two halves based on their projection $t_i$ relative to the median:
\begin{align}
    \mathcal{H}_1 &= \{i : t_i < \text{median}(t)\} \quad \text{(near $t_{\min}$ end)}, \\
    \mathcal{H}_2 &= \{i : t_i \geq \text{median}(t)\} \quad \text{(near $t_{\max}$ end)}.
\end{align}

For each hit, the transverse distance from the principal axis is:
\begin{equation}
    d_\perp^{(i)} = \sqrt{|\vec{r}_i - \bar{\vec{r}}|^2 - t_i^2}.
\end{equation}

The mean transverse spread for each half is:
\begin{equation}
    \sigma_1 = \frac{1}{|\mathcal{H}_1|} \sum_{i \in \mathcal{H}_1} d_\perp^{(i)}, \quad
    \sigma_2 = \frac{1}{|\mathcal{H}_2|} \sum_{i \in \mathcal{H}_2} d_\perp^{(i)}.
\end{equation}

\paragraph{Step 4: Vertex Selection via Transverse Spread Disambiguation}

The vertex is identified as the extremal \emph{hit} at the end with smaller transverse spread:
\begin{equation}
    \vec{v}_{\text{reco}} =
    \begin{cases}
        \vec{r}_{\min} & \text{if } \sigma_1 \leq \sigma_2, \\
        \vec{r}_{\max} & \text{otherwise}.
    \end{cases}
\end{equation}

This algorithm requires no prior knowledge of the true vertex location. Instead, it relies on the universal shower topology: the vertex end has smaller transverse spread ($\sigma \sim p$) because hits converge there, while the shower end has larger spread ($\sigma \sim \alpha L$, where $\alpha$ is the cone angle and $L$ is the shower length). By identifying which half is ``tighter,'' we determine which end is the vertex.

\paragraph{Resolution}

Crucially, we return the \emph{actual extremal hit position} rather than extrapolating the PCA axis beyond the hit distribution. This ensures the vertex resolution is limited by the voxel pitch $p$ rather than accumulated uncertainties from axis fitting.
The expected resolution is $\sigma_v \approx p/\sqrt{12} \approx 0.29$ cm for a 1 cm pitch detector. Our simulations confirm this performance for the 3-View geometry across all orientations.

\begin{figure}[H]
    \centering
    \includegraphics[width=1.0\linewidth]{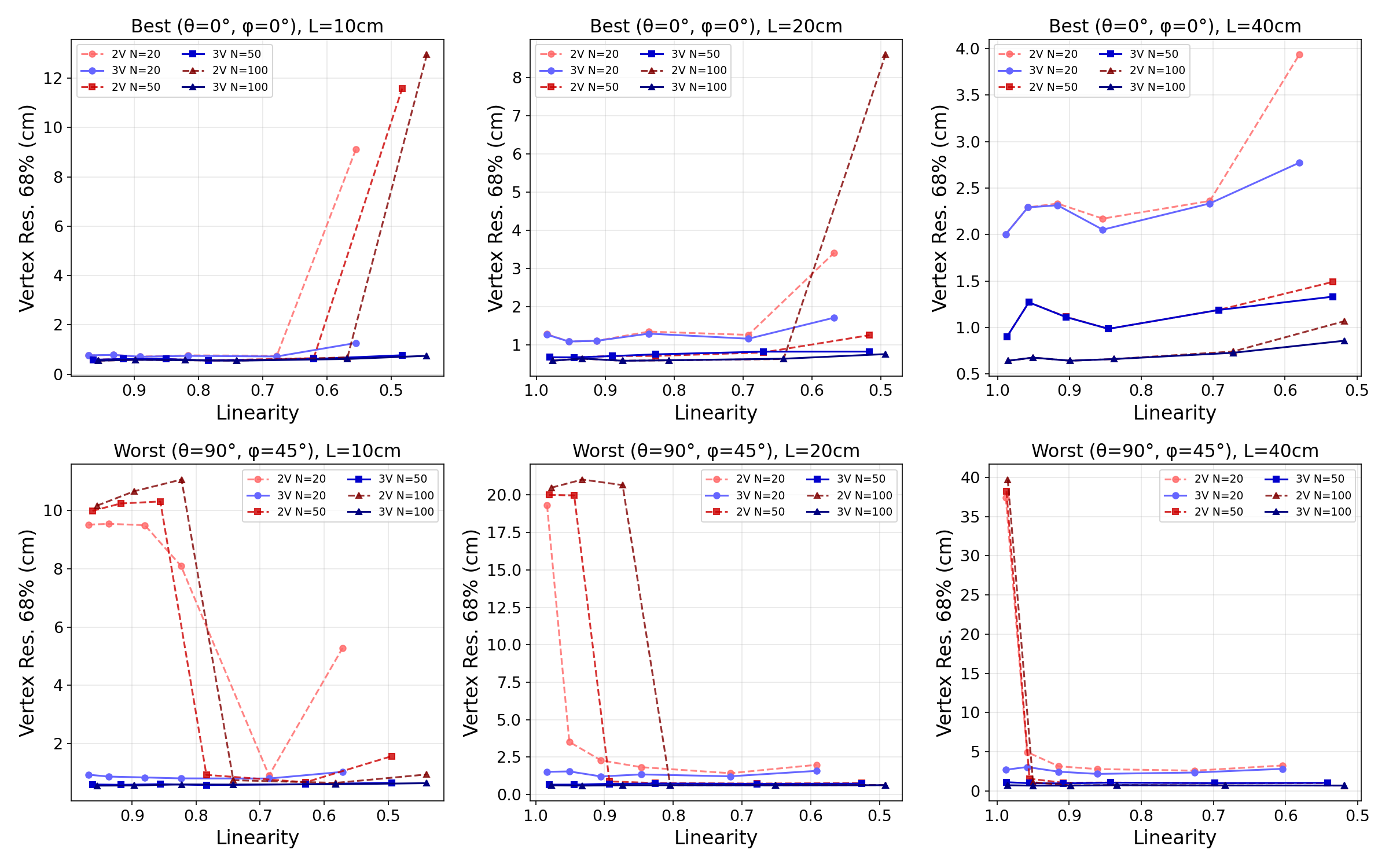}
    \caption{Single-shower vertex resolution (68th percentile, $N=50$ hits) vs linearity. Top row: best case ($\theta = 0^\circ$, $\phi = 0^\circ$). Bottom row: worst case ($\theta = 90^\circ$, $\phi = 45^\circ$). Columns: shower length $L = 10, 20, 40$ cm. Red dashed: 2-View. Blue solid: 3-View.}
    \label{fig:single_vertex}
\end{figure}

However, the 2-View geometry exhibits a strong orientation dependence. At the worst-case diagonal orientation $(\theta=90^\circ, \phi=45^\circ)$, the ghost hits form a dense, symmetric wall that mimics the true track. For narrow cones (high linearity), this symmetry renders the transverse spread disambiguation ineffective, leading to degraded resolution (see Figure \ref{fig:single_vertex}, bottom row). Interestingly, as the cone widens (lower linearity), the ghost distribution disperses, breaking the symmetry and allowing the algorithm to recover the true vertex with improved accuracy.
Using the same cone-shaped clusters, Figure \ref{fig:single_vertex} shows the vertex resolution achieved by both geometries.

\subsection{Vertex Resolution $(\theta, \phi)$ Dependence}

\begin{figure}[H]
    \centering
    \includegraphics[width=1.0\linewidth]{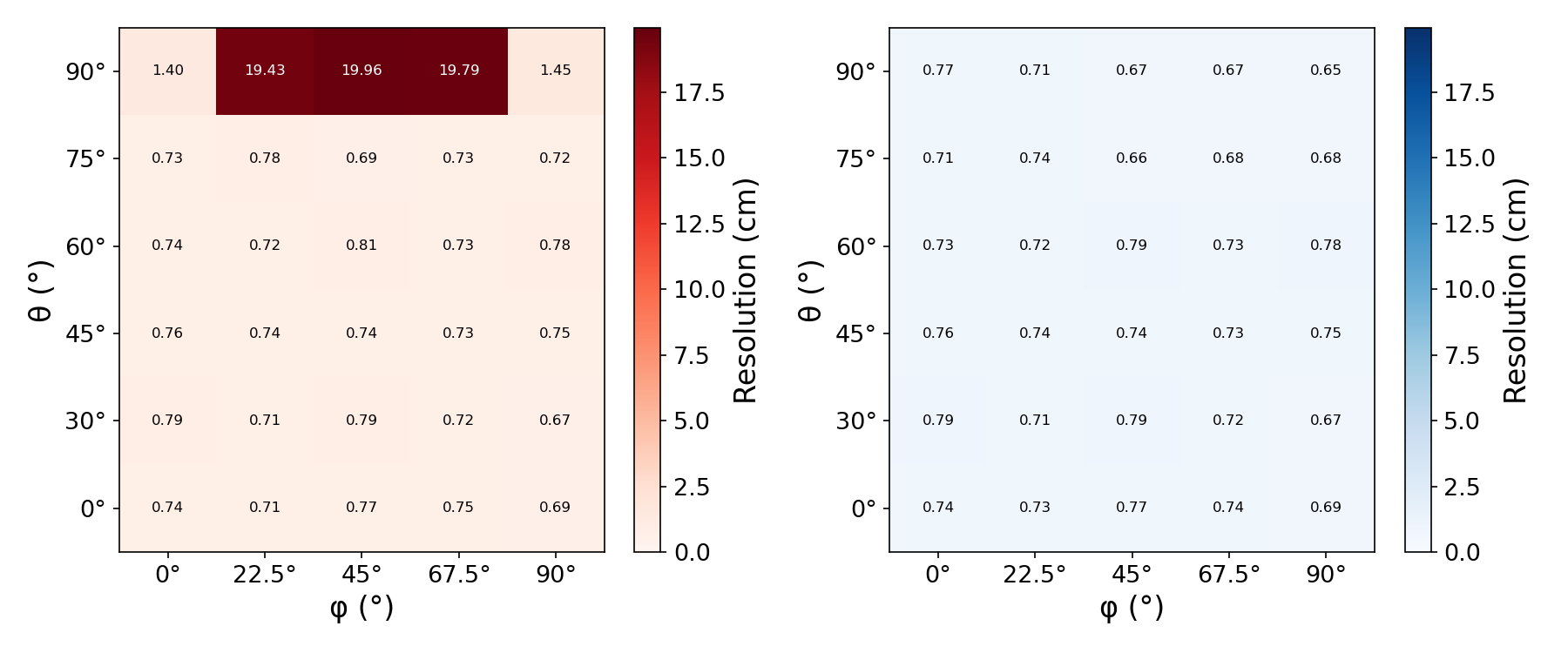}
    \caption{Vertex resolution heatmap ($N=50$ hits, $L=20$ cm) for $\alpha = 5^\circ$ (track-like). Left: 2-View. Right: 3-View. The 2V resolution degrades at $(\theta = 90^\circ, \phi = 45^\circ)$ due to the ``wall of ghosts'' biasing the PCA extrapolation.}
    \label{fig:vertex_heatmap_5}
\end{figure}

\begin{figure}[H]
    \centering
    \includegraphics[width=1.0\linewidth]{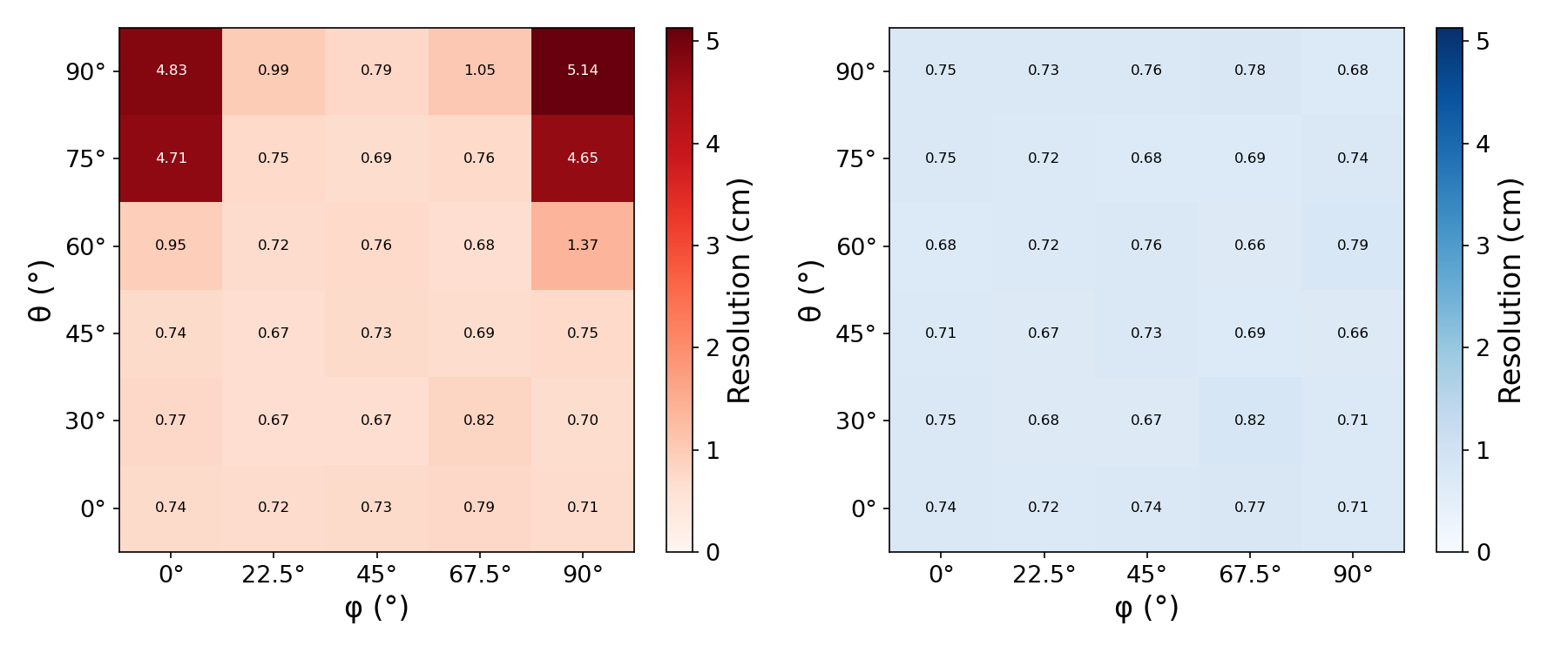}
    \caption{Vertex resolution heatmap ($N=50$ hits, $L=20$ cm) for $\alpha = 30^\circ$ (moderate cone). Left: 2-View. Right: 3-View. The 3V geometry maintains good resolution even at challenging orientations.}
    \label{fig:vertex_heatmap_30}
\end{figure}

Figures~\ref{fig:vertex_heatmap_5} to \ref{fig:vertex_heatmap_30} show the vertex resolution as a function of shower orientation. In each figure, the left panel shows the 2-View result and the right panel shows the 3-View result. The PCA axis extrapolation algorithm achieves sub-centimeter resolution across most orientations, with 3V consistently outperforming 2V.

%==============================================================================
\section{Two-Shower Vertex Resolution}
\label{sec:two_cluster}
%==============================================================================

The canonical application is $\pi^0$ reconstruction via $\pi^0 \to \gamma\gamma$. Both photons convert and produce electromagnetic showers. Reconstructing the common decay vertex helps constrain the $\pi^0$ direction and reject combinatorial backgrounds from random photon pairs.

\subsection{Two-Shower Vertex Finding Algorithm}

When two showers originate from a common vertex, the intersection of their axes provides the vertex position. However, the combined hit pattern must first be clustered into two groups corresponding to each shower. We employ a \emph{hybrid clustering} approach that combines two complementary methods to achieve robust performance across all separation angles.

\paragraph{Method A: Local Density Maxima}

We first identify a seed vertex candidate by finding the hit with the highest local density. For each hit, we count the number of neighbors within a radius $R = 3p$, where $p$ is the voxel pitch. The hit with the maximum neighbor count serves as a density-based vertex estimate, $\vec{v}_\text{density}$. This method is robust against outliers but limited by the voxel quantization.

\paragraph{Method B: Second Principal Component (PC2) Clustering}

For two tracks or showers separated by a moderate angle, the combined hit pattern has two principal directions: PC1 along the average shower direction, and PC2 perpendicular to this, pointing ``across'' the two showers. Projecting hits onto PC2 separates them by shower:
\begin{equation}
    s_i = (\vec{r}_i - \bar{\vec{r}}) \cdot \hat{a}_2,
\end{equation}
where $\hat{a}_2$ is the second eigenvector. Hits are assigned to clusters based on the sign of $s_i$:
\begin{align}
    \mathcal{C}_1^B &= \{i : s_i < 0\}, \\
    \mathcal{C}_2^B &= \{i : s_i \geq 0\}.
\end{align}

\paragraph{Method C: Extremal Distance Clustering.}
For separation angles approaching 90$^\circ$, the PC2 direction may align with one shower rather than across them. In this case, we use \emph{extremal distance clustering}: identify the two hits at the extremes of PC1, then assign each hit to the cluster whose extremal hit is closer:
\begin{align}
    \vec{r}_{\min} &= \vec{r}_{i : t_i = \min(t)}, \quad \vec{r}_{\max} = \vec{r}_{i : t_i = \max(t)}, \\
    d_i^{\min} &= |\vec{r}_i - \vec{r}_{\min}|, \quad d_i^{\max} = |\vec{r}_i - \vec{r}_{\max}|, \\
    \mathcal{C}_1^C &= \{i : d_i^{\min} < d_i^{\max}\}, \\
    \mathcal{C}_2^C &= \{i : d_i^{\min} \geq d_i^{\max}\}.
\end{align}

\paragraph{Hybrid Selection Strategy}

We compute vertex candidates from all three methods. The density-based candidate is always kept. For the clustering methods (B and C), we fit lines to the resulting clusters and compute the intersection point (midpoint of closest approach).

The final vertex selection prioritizes geometric consistency. We check if the intersection candidates fall within the bounding box of the hit distribution (plus a safety margin of $2p$). The first intersection candidate satisfying this containment condition is selected as the vertex. If no intersection candidate is contained within the volume, we select the candidate that minimizes the mean distance to all hits in the event. This hierarchy prioritizes precise line intersections when they are physically meaningful, while falling back to the robust density estimate or mean-distance minimizer in pathological cases.

\paragraph{Line Fitting and Closest Approach}

For each cluster, we fit a line using PCA:
\begin{equation}
    \text{Line}_k: \vec{P}_k + t \hat{d}_k,
\end{equation}
where $\vec{P}_k$ is the centroid and $\hat{d}_k$ is the principal direction of cluster $k$.

The vertex is defined as the midpoint of the line of closest approach between the two lines. For lines $\vec{L}_1(t) = \vec{P}_1 + t\hat{d}_1$ and $\vec{L}_2(s) = \vec{P}_2 + s\hat{d}_2$, the parameters at closest approach satisfy:
\begin{align}
    t^* &= \frac{(\hat{d}_1 \cdot \hat{d}_2)(\hat{d}_2 \cdot \vec{w}_0) - (\hat{d}_2 \cdot \hat{d}_2)(\hat{d}_1 \cdot \vec{w}_0)}{(\hat{d}_1 \cdot \hat{d}_1)(\hat{d}_2 \cdot \hat{d}_2) - (\hat{d}_1 \cdot \hat{d}_2)^2}, \\
    s^* &= \frac{(\hat{d}_1 \cdot \hat{d}_1)(\hat{d}_2 \cdot \vec{w}_0) - (\hat{d}_1 \cdot \hat{d}_2)(\hat{d}_1 \cdot \vec{w}_0)}{(\hat{d}_1 \cdot \hat{d}_1)(\hat{d}_2 \cdot \hat{d}_2) - (\hat{d}_1 \cdot \hat{d}_2)^2},
\end{align}
where $\vec{w}_0 = \vec{P}_1 - \vec{P}_2$. The reconstructed vertex is:
\begin{equation}
    \vec{v}_{\text{reco}} = \frac{1}{2}\left(\vec{L}_1(t^*) + \vec{L}_2(s^*)\right).
\end{equation}

This algorithm achieves sub-centimeter resolution across separation angles from 10$^\circ$ to 90$^\circ$, with typical 68th percentile errors of 0.1 to 0.8 cm for a 1 cm pitch detector.

\subsection{Results}

\begin{figure}[H]
    \centering
    \includegraphics[width=0.85\linewidth]{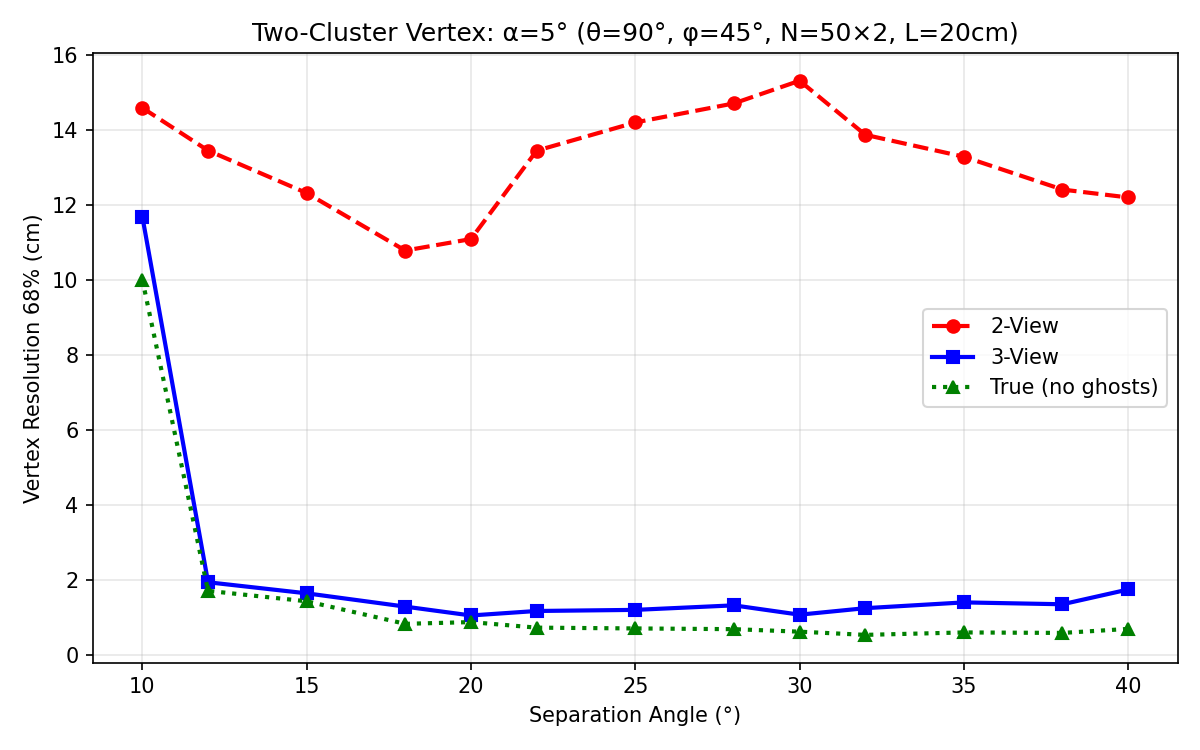}
    \caption{Two-shower vertex resolution for $\alpha = 5^\circ$ (track-like showers) at worst-case orientation ($\theta = 90^\circ$, $\phi = 45^\circ$). $N = 50$ hits per shower, $L = 20$ cm. Red dashed: 2-View ($\sim$11 to 15 cm). Blue solid: 3-View ($\sim$1 to 2 cm). Green dotted: true hits (no ghosts, $\sim$0.5 to 1 cm). The 3V advantage is approximately 10$\times$ due to ghost hit suppression.}
    \label{fig:two_vertex_5}
\end{figure}

\begin{figure}[H]
    \centering
    \includegraphics[width=0.85\linewidth]{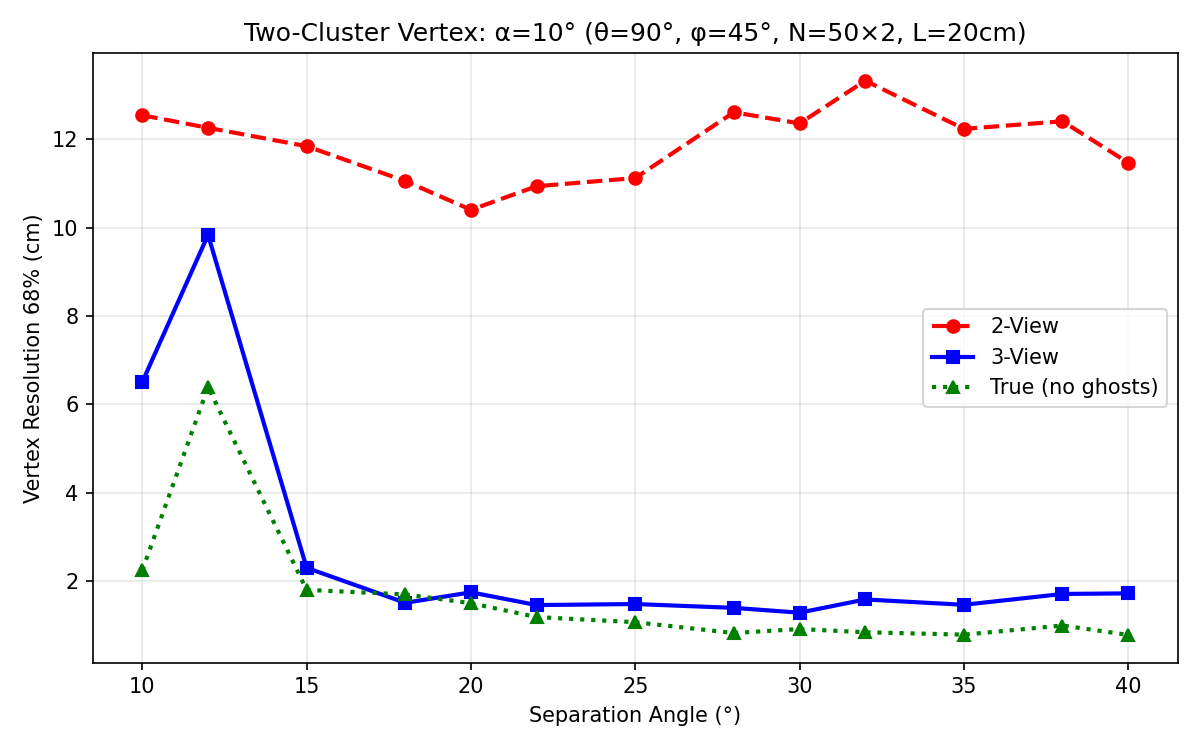}
    \caption{Two-shower vertex resolution for $\alpha = 10^\circ$ at worst-case orientation. Similar to $\alpha = 5^\circ$, with 2V resolution degraded by ghost contamination.}
    \label{fig:two_vertex_10}
\end{figure}

\begin{figure}[H]
    \centering
    \includegraphics[width=0.85\linewidth]{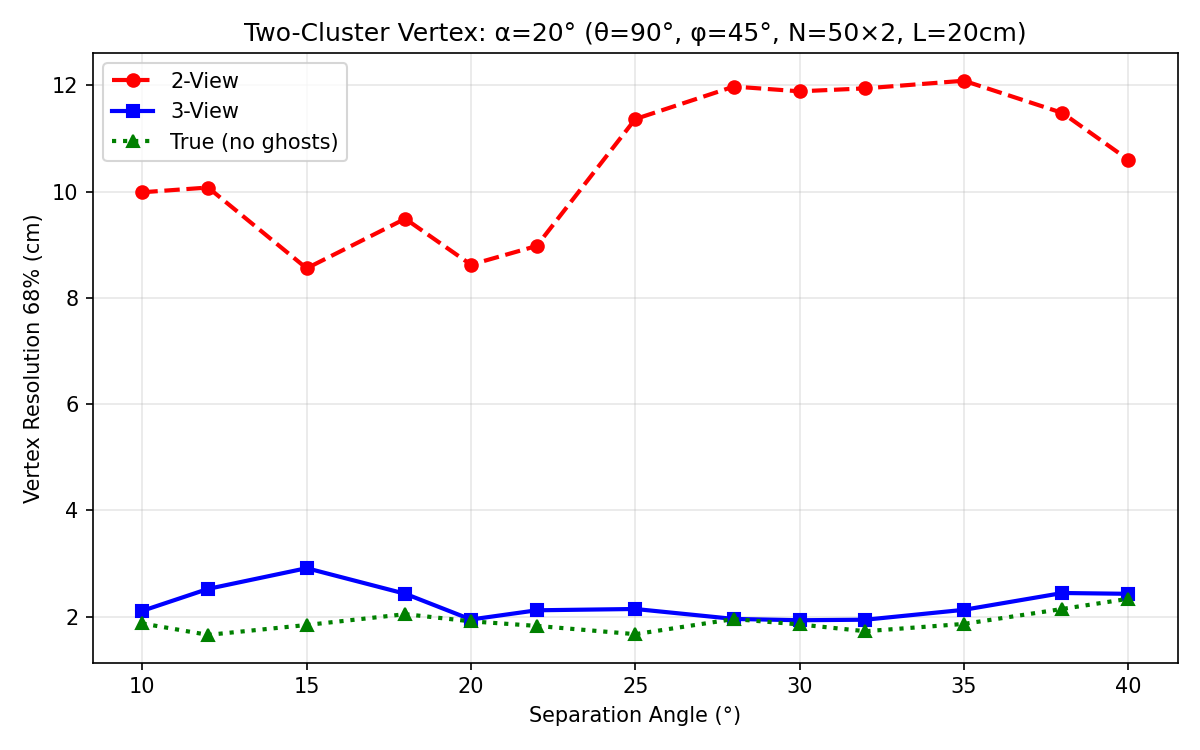}
    \caption{Two-shower vertex resolution for $\alpha = 20^\circ$ at worst-case orientation. Increasing shower opening reduces track linearity, slightly degrading vertex resolution for all methods.}
    \label{fig:two_vertex_20}
\end{figure}

We generate two cone-shaped showers originating from a common vertex. Each shower has $N = 50$ hits and length $L = 20$ cm. Both showers are oriented at the worst-case direction ($\theta = 90^\circ$, $\phi = 45^\circ$) to maximize ghost hit contamination, with varying separation angles between 10$^\circ$ and 40$^\circ$.
Figures~\ref{fig:two_vertex_5} to \ref{fig:two_vertex_20} show the vertex resolution as a function of separation angle for three cone opening angles: $\alpha = 5^\circ$ (track-like), $10^\circ$, and $20^\circ$ (shower-like).

One can observe a few things from the figures:
\begin{itemize}
    \item At worst-case orientation, 2V vertex resolution is 5 to 10$\times$ worse than 3V due to ghost hit contamination biasing cluster separation.
    \item 3V resolution closely tracks the ghost-free case, confirming effective ghost suppression.
    \item Resolution degrades with increasing cone opening ($\alpha$) as showers become less linear.
    \item The hybrid clustering algorithm maintains robustness across separation angles from 10$^\circ$ to 40$^\circ$.
\end{itemize}

%==============================================================================
\section{Energy-Dependent Performance for Particles through Matter}
\label{sec:energy}
%==============================================================================

Real particle showers exhibit energy-dependent characteristics. Electrons initiate electromagnetic cascades through bremsstrahlung and pair production, creating spatially extended hit patterns that scale with energy. Muons undergo multiple Coulomb scattering that deflects them from straight trajectories. Understanding how 2V/3V performance varies with energy is essential for experiments spanning wide kinematic ranges. We emphasize that the simulations in this section are \emph{not} based on a full Geant4 Monte Carlo, but instead employ parametric models: the Bethe-Bloch formula for ionization energy loss, the Highland approximation for multiple Coulomb scattering, and the Moli\`ere parameterization for electromagnetic shower lateral spread. This allows us to study the readout geometry effect in isolation, without the complications of shower-by-shower fluctuations that a full Geant4 simulation would introduce.

We study three scintillator materials (Table~\ref{tab:material}) \cite{PDG}: plastic scintillator (low-Z, long radiation length), BGO, and lead tungstate (high-Z, short radiation length). For each material, track length and hit count are derived from particle physics rather than fixed parameters.

\begin{table}[h]
\centering
\caption{Material properties for three scintillator types used in the energy-dependent study.}
\label{tab:material}
\begin{tabular}{lccc}
\hline
\textbf{Property} & \textbf{plastic} & \textbf{BGO} & \textbf{PbWO$_4$} \\
\hline
Radiation length, $X_0$ (cm) & 42.4 & 1.12 & 0.89 \\
Density, $\rho$ (g/cm$^3$) & 1.03 & 7.13 & 8.28 \\
Critical energy, $E_c$ (MeV) & 80 & 10.1 & 9.6 \\
Mean ionization potential, $I$ (eV) & 68.7 & 534 & 600 \\
\hline
\end{tabular}
\end{table}

\paragraph{Multiple Scattering (Highland)}

For a particle of momentum $p$ (MeV/c) and velocity $\beta c$ traversing material of thickness $x$, we use the Highland formula \cite{Highland_1975, Lynch_Dahl_1991}:

\begin{equation}
    \theta_{\text{RMS}} = \frac{13.6}{\beta p} \sqrt{\frac{x}{X_0}} \left[1 + 0.038 \ln\left(\frac{x}{X_0}\right)\right]
\end{equation}

\paragraph{Electron Shower Opening}

Electromagnetic shower angular spread is modeled using the Moli\`ere angle \cite{Rossi_Greisen_1941}: $\theta_M \approx 21\text{ MeV}/E$.

\subsection{Energy vs Angular Resolution at Three Orientations}

We simulate muons and electrons from 100 MeV to 10 GeV in three materials: plastic scintillator (low-Z), BGO, and lead tungstate (high-Z). Track length and hit count are derived from physics: for muons, range $L = E / (dE/dx)$, where $dE/dx$ is computed from the Bethe-Bloch formula; for EM showers, length $L \approx 2 X_0 (1 + \ln(E/E_c))$, with number of hits scaling with $E/E_c$.

\begin{figure}[H]
    \centering
    \includegraphics[width=1.0\linewidth]{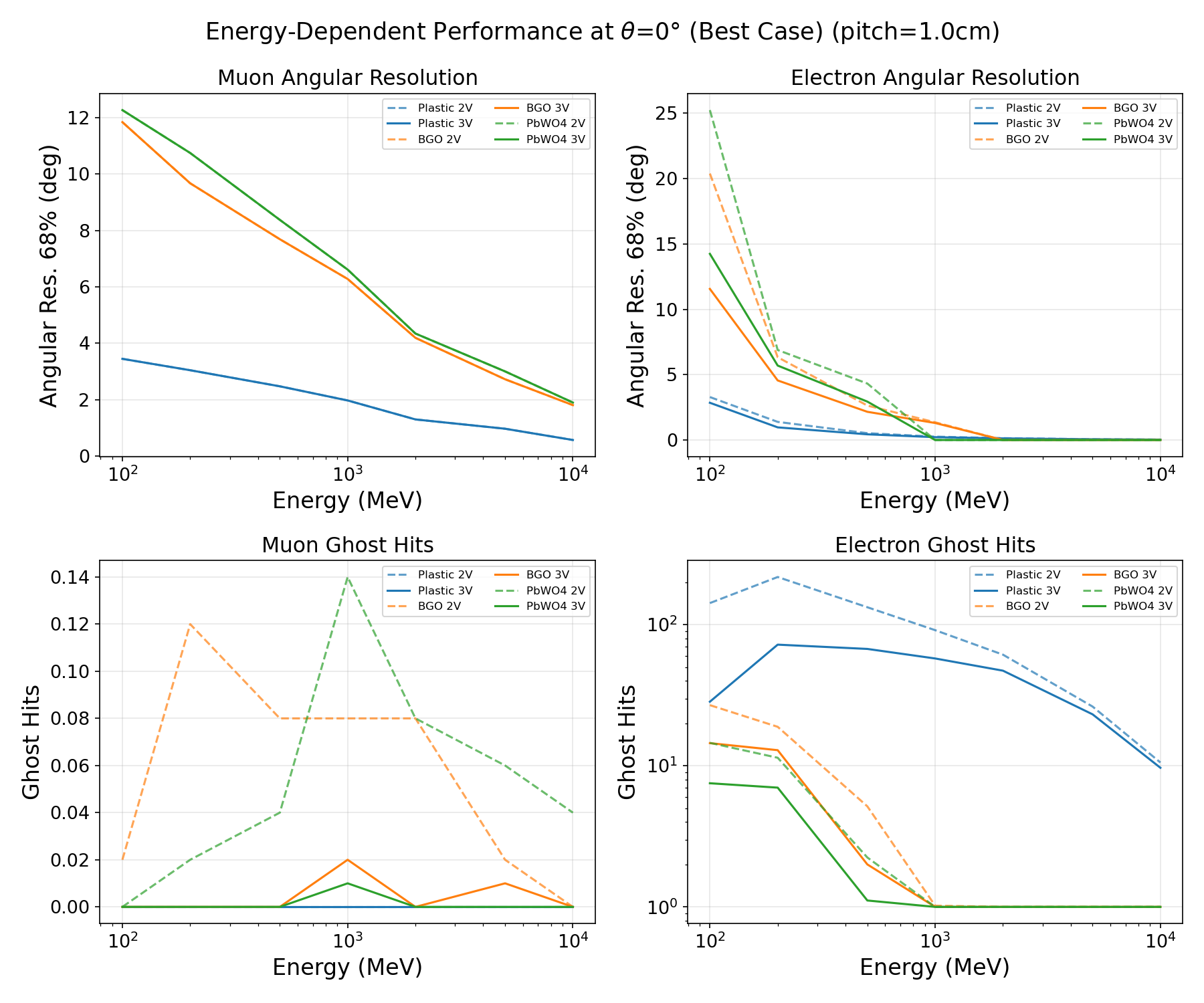}
    \caption{Energy-dependent performance at $\theta = 0^\circ$ (best case, along Z-axis). Track length and hit count derived from physics. Top row: angular resolution (68\%). Bottom row: ghost hits (log scale). Solid: 3V. Dashed: 2V. Blue: plastic. Orange: BGO. Green: PbWO$_4$. At this orientation, 2V and 3V have similar performance.}
    \label{fig:energy_theta0}
\end{figure}

\begin{figure}[H]
    \centering
    \includegraphics[width=1.0\linewidth]{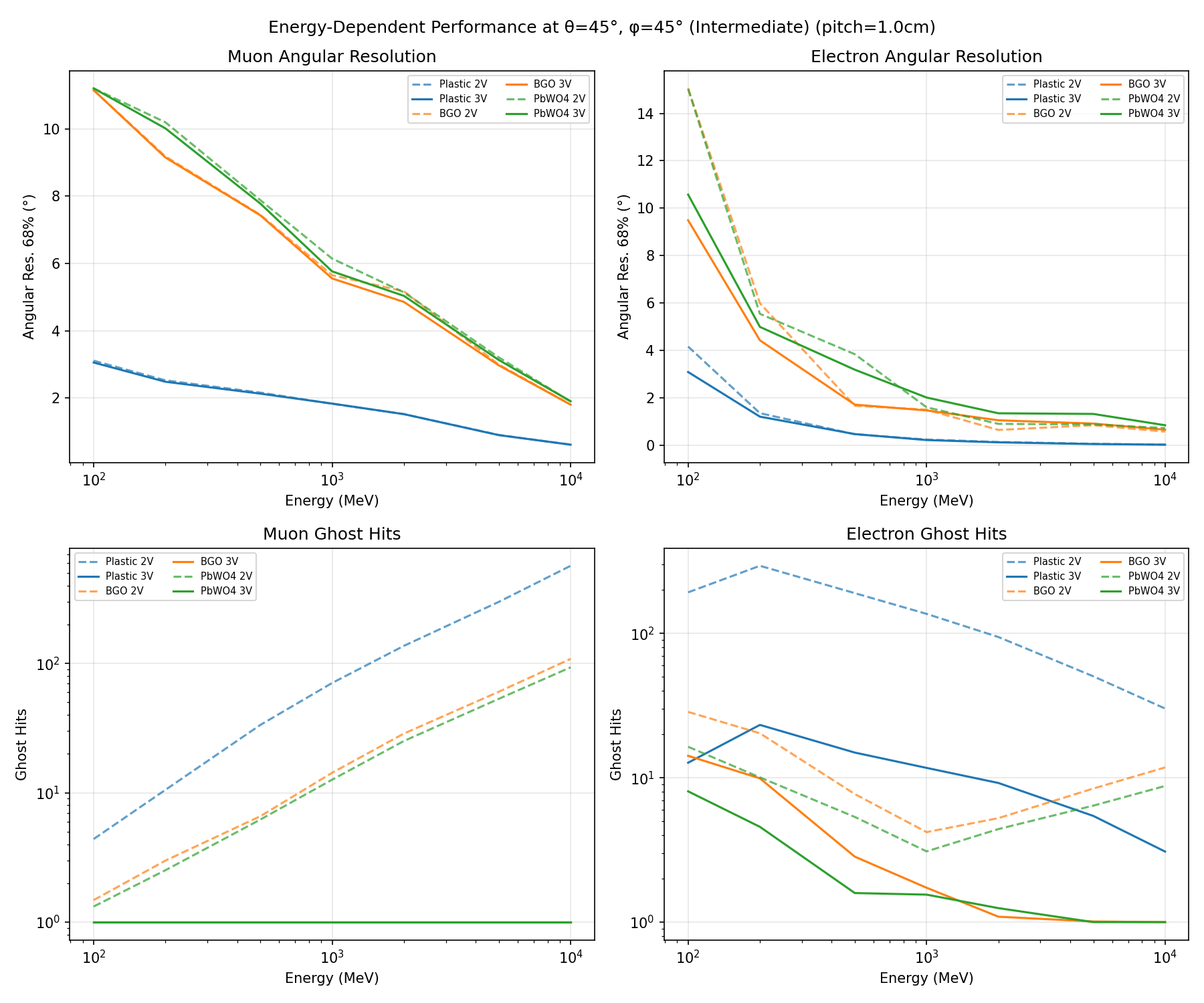}
    \caption{Energy-dependent performance at $\theta = 45^\circ$, $\phi = 45^\circ$ (intermediate case). At this orientation, moderate ghost contamination begins to appear in 2V. Multiple scattering limits low-energy resolution for all materials.}
    \label{fig:energy_theta45}
\end{figure}

\begin{figure}[H]
    \centering
    \includegraphics[width=1.0\linewidth]{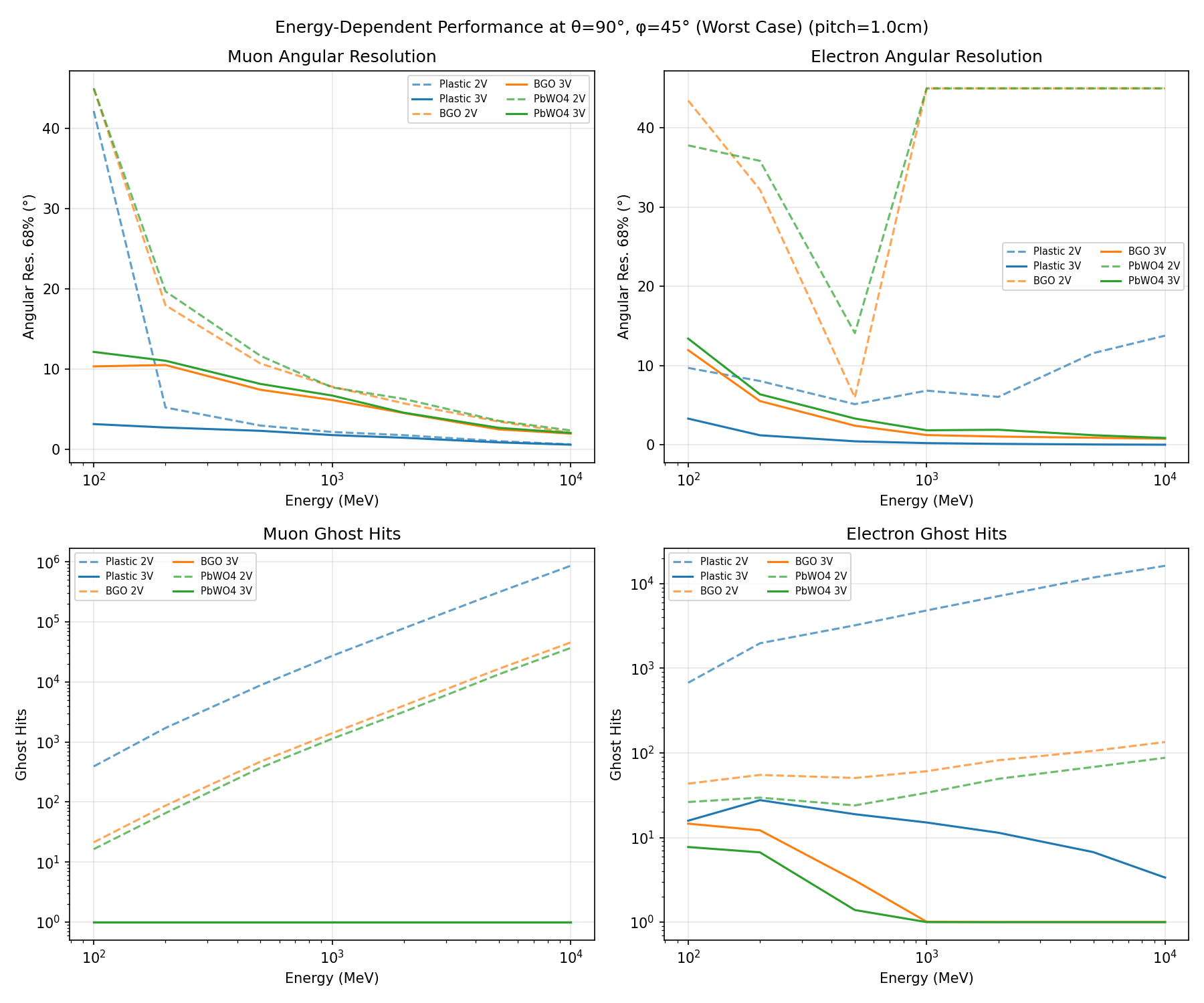}
    \caption{Energy-dependent performance at $\theta = 90^\circ$, $\phi = 45^\circ$ (worst case). 2V ghost hits increase dramatically (100 to 1000s), causing angular resolution to degrade to $\sim$45$^\circ$ at low energies. 3V maintains resolution below 10$^\circ$ across all energies.}
    \label{fig:energy_theta90}
\end{figure}

We simulate three materials: plastic scintillator, characterized by low density ($\rho \approx 1$ g/cm$^3$) and a long radiation length ($X_0 \approx 42$ cm); BGO, with high density ($\rho \approx 7.1$ g/cm$^3$) and a short radiation length ($X_0 \approx 1.1$ cm); and PbWO$_4$, which has very high density ($\rho \approx 8.3$ g/cm$^3$) and a very short radiation length ($X_0 \approx 0.9$ cm).
Figures~\ref{fig:energy_theta0}, \ref{fig:energy_theta45}, and \ref{fig:energy_theta90} show angular resolution and ghost hits as a function of energy at three orientations. In each figure, dashed lines represent 2-View and solid lines represent 3-View results.

Plastic scintillator, due to its low density, produces long tracks ($L \sim 200$ cm at 1 GeV), which leads to massive ghost contamination in the 2-View geometry. In contrast, high-Z materials like BGO and PbWO$_4$ result in shorter tracks but exhibit more severe multiple scattering per unit length. Consequently, the 3-View advantage is most pronounced at worst-case orientations and lower energies.

\paragraph{Physical Interpretation of Features}

Two specific features in the worst-case orientation (Figure~\ref{fig:energy_theta90}) warrant explanation:
\begin{enumerate}
    \item \textbf{Muon Resolution Rise (plastic 2V)}

    Above 1 GeV, the angular resolution for muons in plastic scintillator degrades significantly in 2-View at the worst-case orientation. This is due to the long track length ($L \approx 200$ cm) in low-density material. At the diagonal orientation, a 200-hit track generates $\sim 40,000$ ghost hits, creating a dense plane of false signals that overwhelms the true track and biases the PCA axis. This degradation is specific to the 2-View geometry at diagonal orientations; the 3-View geometry maintains stable sub-degree resolution across all energies at this orientation (see Figure~\ref{fig:energy_theta90}, top-left panel).

    Crucially, the resolution does not simply scale with ghost count. At low energies ($<$1 GeV), plastic 2V performs \emph{better} than high-Z materials despite having \emph{more} ghosts. This is because plastic's low density allows for much longer tracks (larger lever arm) and less multiple scattering, which dominates the resolution budget. The ghost penalty only becomes catastrophic above 1 GeV when the combinatorial background saturates, overwhelming the lever arm advantage.

    In contrast, high-Z materials (BGO, PbWO$_4$) perform well at high energy. Their high density results in much higher $dE/dx$, keeping tracks physically short ($\sim$20 to 30 cm) even at 10 GeV. This limits the ghost count to manageable levels ($\sim$900 ghosts vs 40,000 for plastic). Combined with the natural suppression of multiple scattering at high momentum ($\theta_{\text{MS}} \propto 1/p$), this allows high-Z materials to achieve excellent angular resolution in the high-energy limit, avoiding the ``ghost catastrophe'' that affects plastic.
    \item \textbf{Electron Resolution Dip (High-Z 2V)}

    For BGO and PbWO$_4$, the 2-View electron resolution shows a local minimum (improvement) around 500 to 1000 MeV. This reflects a competition between shower elongation and ghost density. At low energies ($<$500 MeV), showers are short and isotropic (``blob-like''), making direction finding difficult. As energy increases, the shower elongates, improving the PCA definition. However, at very high energies ($>$2 GeV), the hit density increases to the point where the quadratic ghost growth again degrades performance. This sweet spot does not appear for plastic scintillator due to its much longer radiation length ($X_0 = 42$ cm), which produces sparser showers.
\end{enumerate}

\subsection{Angular Resolution Heatmaps at Fixed Energies}

\begin{figure}[H]
    \centering
    \includegraphics[width=1.0\linewidth]{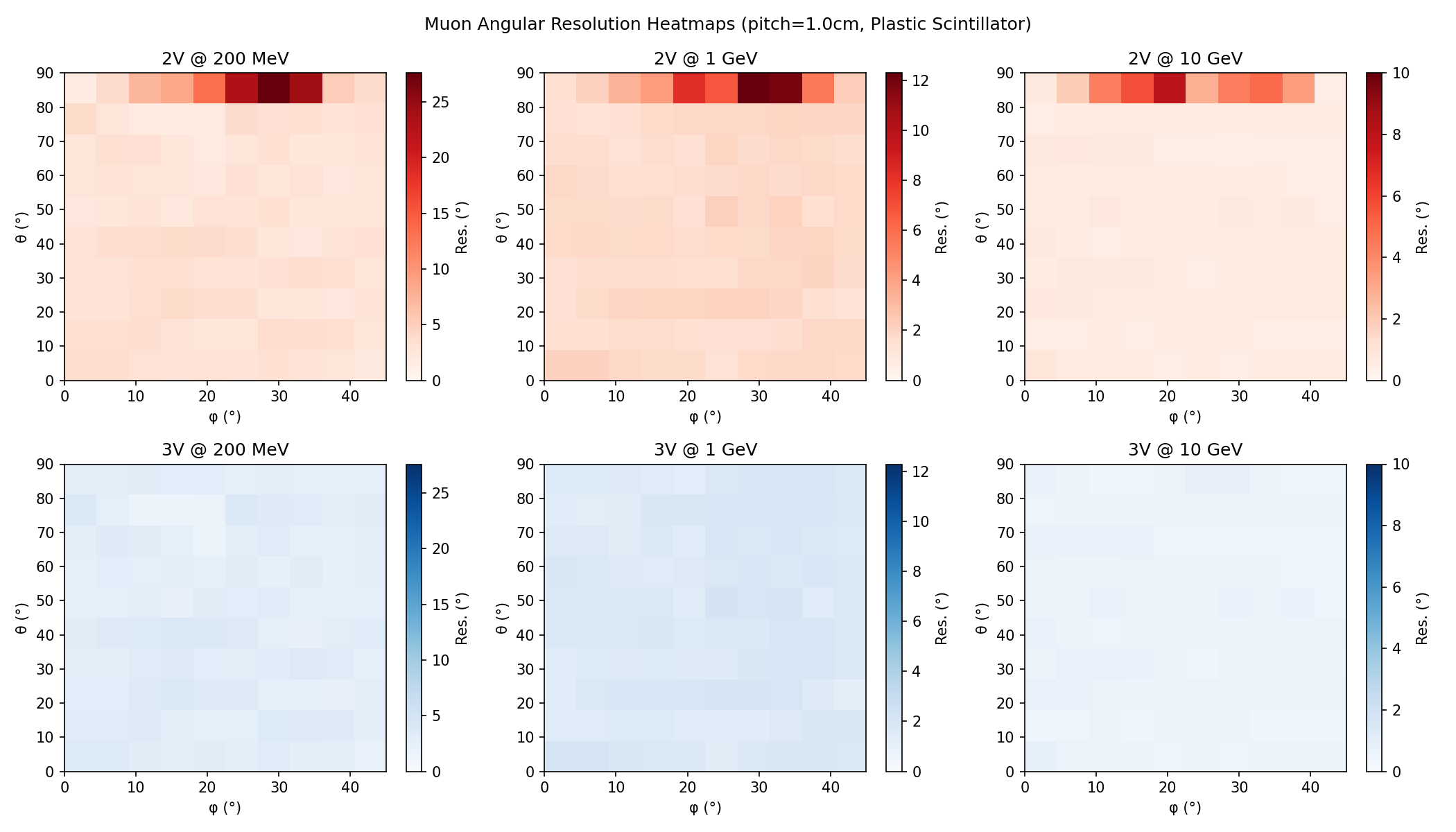}
    \caption{Angular resolution heatmaps for muons in plastic scintillator at fixed energies. Top row: 2V. Bottom row: 3V. Left to right: 200 MeV, 1 GeV, 10 GeV. The 2V resolution shows strong orientation dependence at $(\theta, \phi) \approx (90^\circ, 45^\circ)$ (worst case), while 3V maintains uniform performance across all orientations. At 10~GeV, the 2V worst-case resolution exceeds $40^\circ$, demonstrating the ghost catastrophe for long tracks at diagonal orientations.}
    \label{fig:energy_heatmaps_muon}
\end{figure}

\begin{figure}[H]
    \centering
    \includegraphics[width=1.0\linewidth]{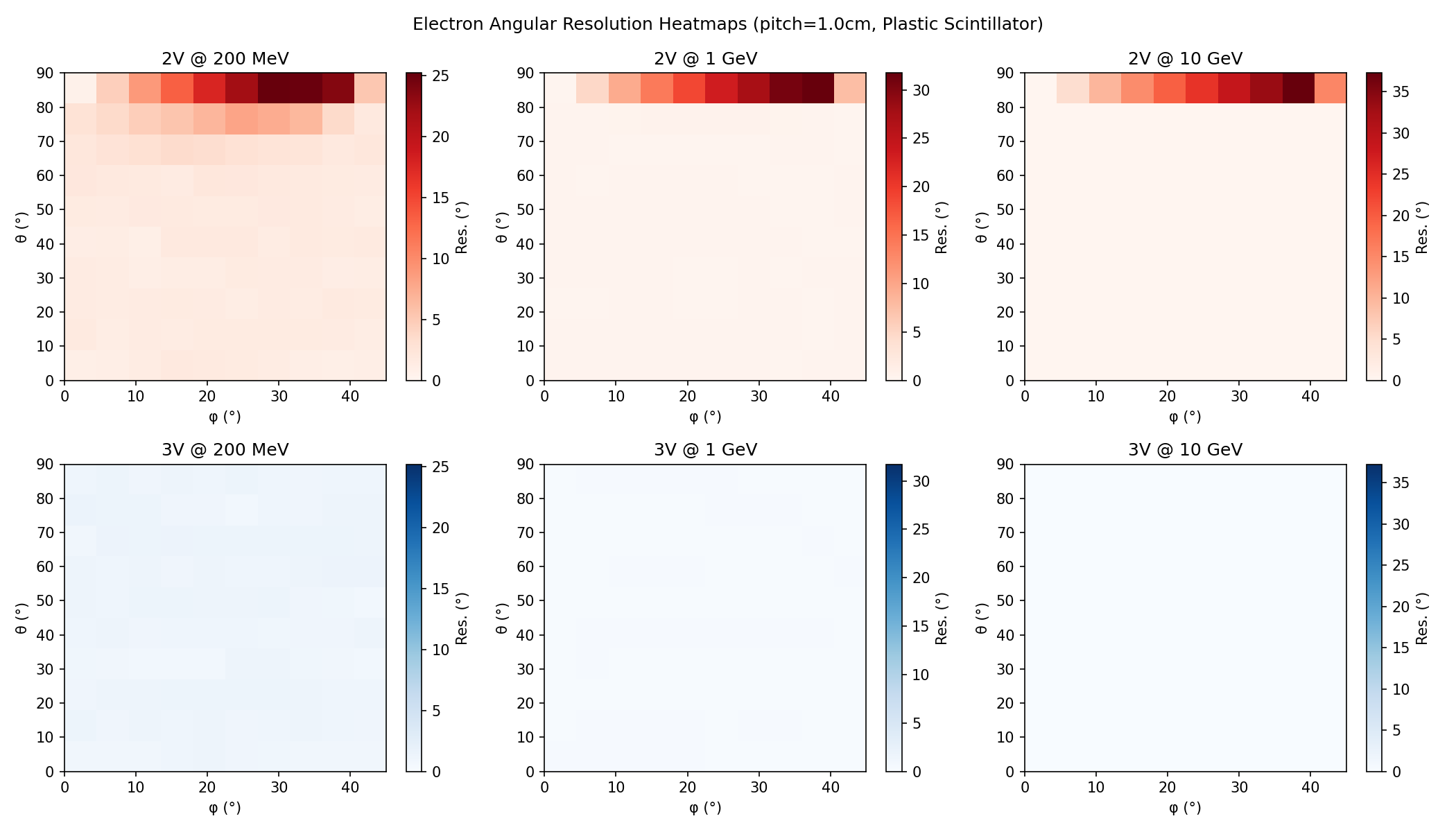}
    \caption{Angular resolution heatmaps for electrons in plastic scintillator at fixed energies. Top row: 2V. Bottom row: 3V. Left to right: 200 MeV, 1 GeV, 10 GeV. Electrons exhibit broader angular spread due to electromagnetic shower development, but the 3V advantage persists across all energies and orientations.}
    \label{fig:energy_heatmaps_electron}
\end{figure}

Figures~\ref{fig:energy_heatmaps_muon} and \ref{fig:energy_heatmaps_electron} show angular resolution heatmaps for muons and electrons at 200 MeV, 1 GeV, and 10 GeV. In each figure, the top row shows the 2-View result and the bottom row shows the 3-View result.

Figure~\ref{fig:energy_heatmaps_muon} shows the orientation dependence clearly: as the muon energy increases from 200~MeV to 10~GeV, the 2V angular resolution at the worst-case diagonal orientation degrades from $\sim$10$^\circ$ to $>$40$^\circ$, while 3V stays below 5$^\circ$ everywhere.

In Figures~\ref{fig:energy_heatmaps_muon} and \ref{fig:energy_heatmaps_electron}, the worst 2-View resolution is not exactly at $\phi = 45^\circ$ but at nearby angles. As discussed in Section~\ref{sec:angular}, at exactly $\phi = 45^\circ$ the ghost distribution is symmetric about the track axis and biases the PCA fit less than the asymmetric ghost pattern at neighboring angles. The same effect shows up in the vertex heatmaps in Appendix~\ref{app:fine_pitch}.

%==============================================================================
\section{Conclusion}
\label{sec:conclusion}
%==============================================================================

Fine-grained scintillator detectors represent a transformative leap in our ability to image particle interactions with high fidelity. This study has systematically quantified the performance advantages of the 3-View readout geometry over the traditional 2-View approach. By introducing a third orthogonal projection, the 3-View design fundamentally breaks the combinatorial ambiguities that plague 2-View systems, reducing ghost hit contamination by up to 90\% in high-multiplicity environments.

Our benchmarks demonstrate that this topological clarity translates directly into superior reconstruction performance. The 3-View geometry maintains robust angular and vertex resolution across all particle orientations, eliminating the blind spots at diagonal angles that severely degrade 2-View performance. This isotropy is particularly critical for next-generation neutrino experiments, where the accurate reconstruction of high-angle particles is essential for reducing systematic uncertainties in cross-section measurements. Furthermore, the ability to resolve multi-shower vertices with sub-centimeter precision opens new windows for background rejection in rare decay searches and precision calorimetry at future colliders such as EIC and FCC.

Of course, the 3-View readout comes at a cost. Adding a third fiber plane increases the total channel count by roughly 50\%, with corresponding increases in MPPC and front-end electronics expenses. Mechanically, threading a third set of orthogonal fibers through each scintillator cube demands tighter tolerances and more complex assembly. The DAQ must also synchronize three views instead of two. On the other hand, the cleaner hit reconstruction reduces the burden on offline ghost-rejection algorithms, simplifying the analysis chain and lowering systematic uncertainties. For $4\pi$ near detectors in neutrino oscillation experiments, where orientation-independent performance matters most, we believe the 3-View geometry represents a worthwhile investment.

Looking forward, the adoption of 3D-projection technology is poised to become the standard for high-granularity detectors. As beam intensities increase and physics goals demand ever-higher precision, the limitations of 2D projective readout become a bottleneck. The 3-View architecture offers a cost-effective pathway to true 3D reconstruction, delivering the voxel-level granularity required to disentangle complex event topologies without the prohibitive channel count of fully pixelated readouts. These results provide a quantitative foundation for the design and optimization of future detectors, ensuring they can fully exploit the physics potential of the coming decade.

\bibliographystyle{unsrt}
\bibliography{references}

\newpage
\appendix

\section{Fine Pitch Studies (0.5 cm)}
\label{app:fine_pitch}

This appendix presents a complete replication of all studies from the main paper using a finer voxel pitch of $p = 0.5$ cm, compared to the baseline $p = 1.0$ cm. The purpose is to demonstrate that the 2V/3V performance comparison scales predictably with detector granularity, and to quantify the resolution improvements achievable with finer segmentation.

\subsection{Scaling Expectations}

The ghost hit phenomenon is purely topological---it arises from the combinatorial ambiguity of fiber crossings and is independent of absolute scale. Therefore, ghost hit \emph{counts} remain unchanged when scaling the detector pitch. However, angular and vertex resolution improve approximately linearly with finer pitch:
\begin{equation}
    \sigma_{\text{fine}} \approx \sigma_{\text{baseline}} \times \frac{p_{\text{fine}}}{p_{\text{baseline}}} = \sigma_{\text{baseline}} \times 0.5
\end{equation}

This scaling holds because resolution is fundamentally limited by the voxel size, not by algorithmic uncertainties.

\subsection{Angular Resolution vs Cone Opening}

\begin{figure}[H]
    \centering
    \includegraphics[width=1.0\linewidth]{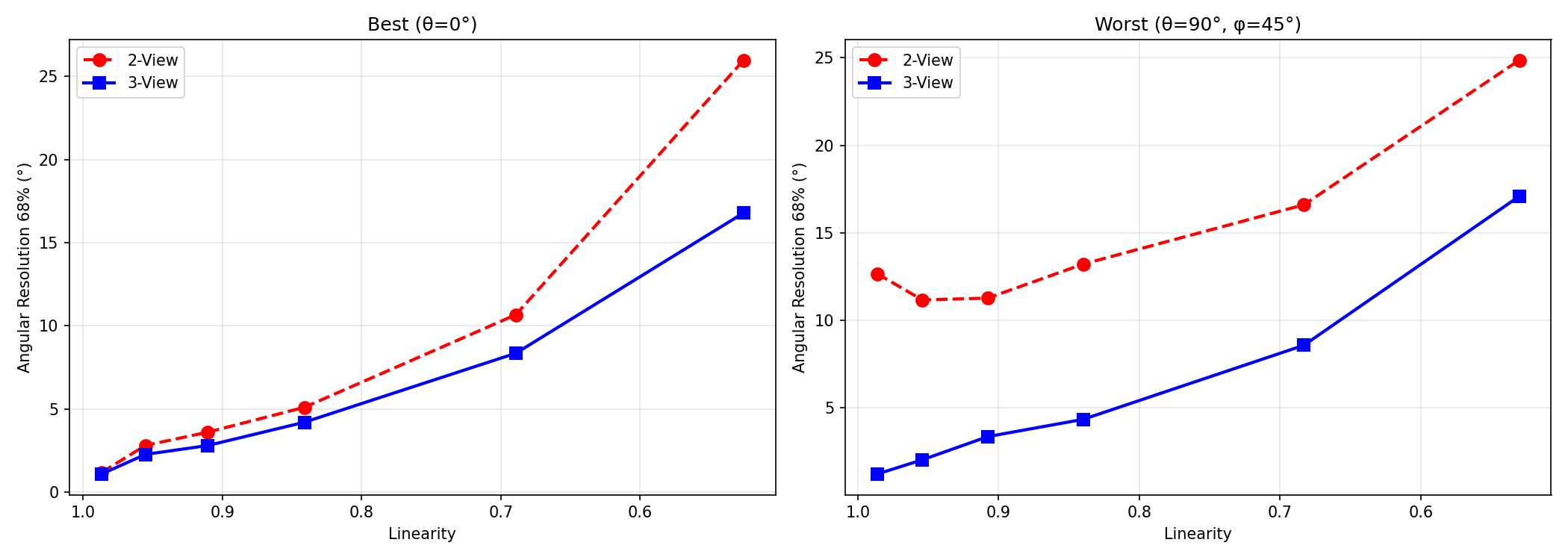}
    \caption{Angular resolution vs cone opening angle at 0.5 cm pitch ($N=50$ hits, $L=20$ cm). Red dashed: 2V. Blue solid: 3V. Resolution improves by approximately 2$\times$ compared to 1.0 cm pitch, while the 3V advantage (factor of 2--3$\times$ at small cone angles) persists.}
    \label{fig:angular_2v3v_5mm}
\end{figure}

Figure~\ref{fig:angular_2v3v_5mm} shows the angular resolution as a function of cone opening angle at 0.5 cm pitch, analogous to the main paper's baseline study.

\begin{figure}[H]
    \centering
    \includegraphics[width=1.0\linewidth]{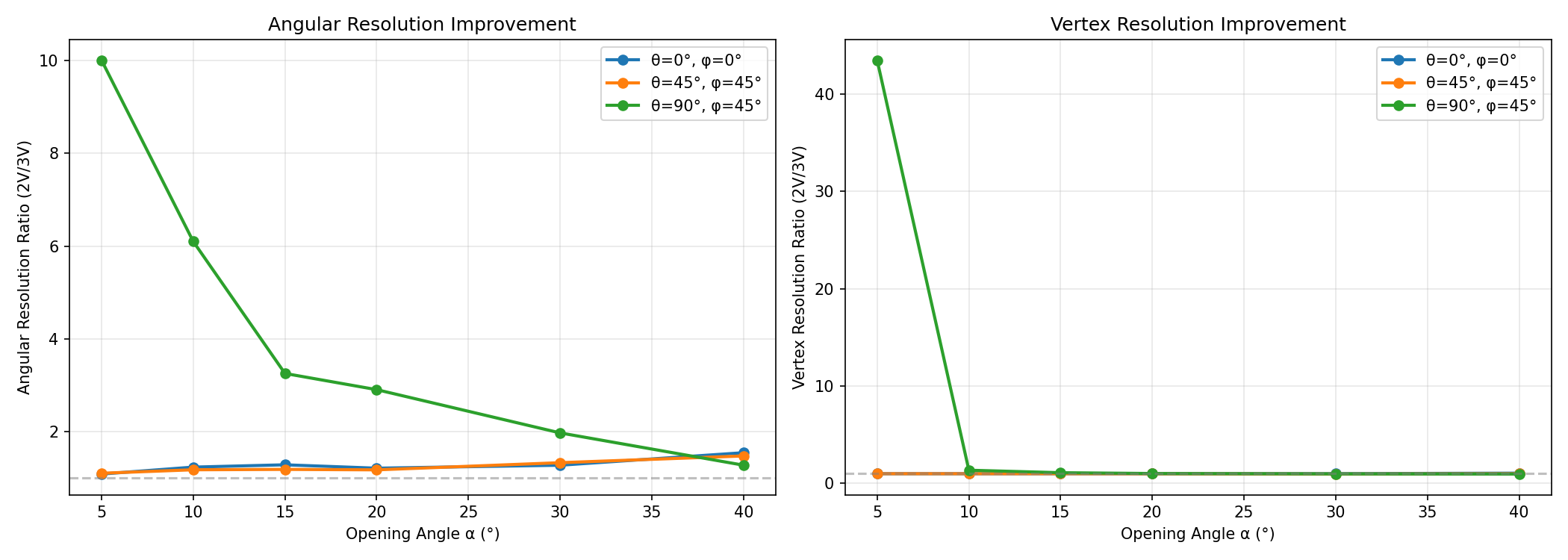}
    \caption{3V/2V improvement ratio at 0.5 cm pitch. The relative advantage remains consistent with the 1.0 cm baseline.}
    \label{fig:improvement_5mm}
\end{figure}

\subsection{Angular Resolution Heatmaps}

\begin{figure}[H]
    \centering
    \includegraphics[width=1.0\linewidth]{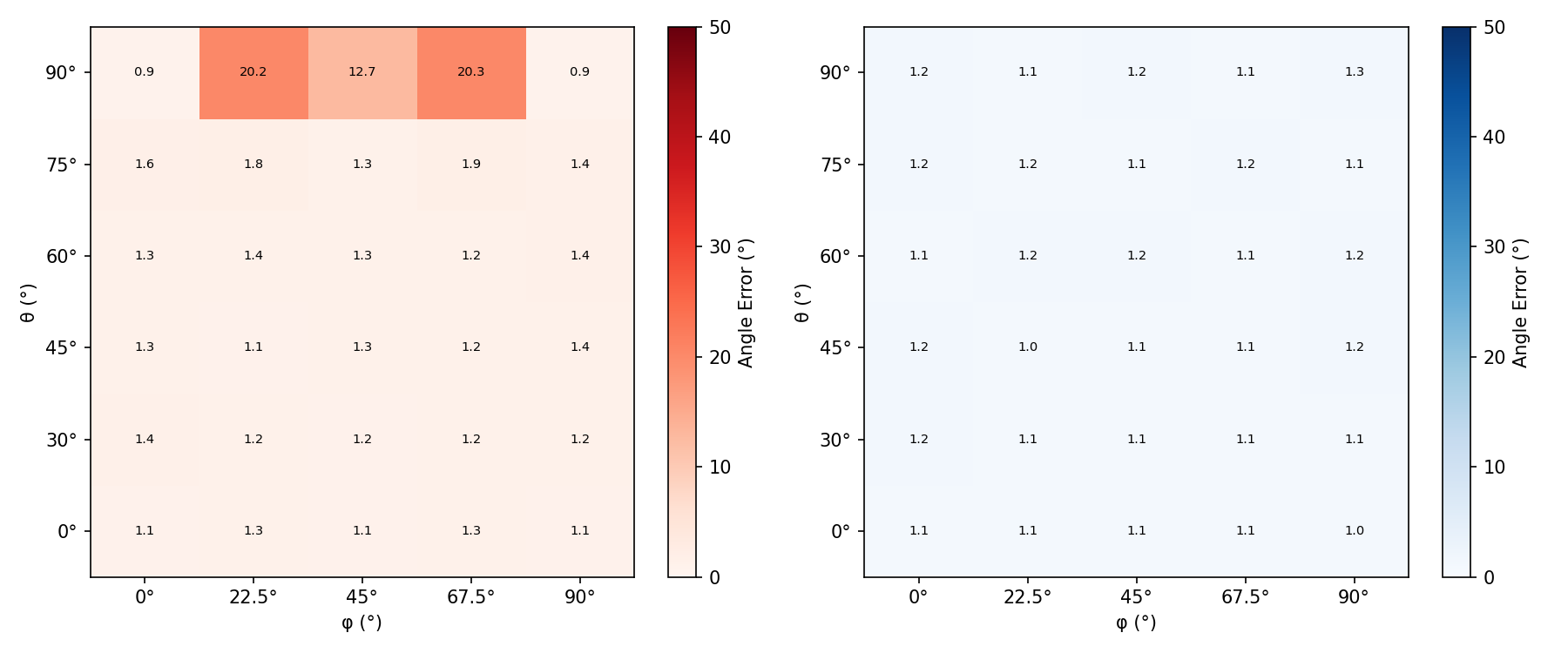}
    \caption{Angular resolution heatmap at 0.5 cm pitch for $\alpha = 5^\circ$ (track-like). The characteristic 2V degradation at $(\theta, \phi) \approx (90^\circ, 45^\circ)$ persists.}
    \label{fig:angular_heatmap_5_5mm}
\end{figure}

\begin{figure}[H]
    \centering
    \includegraphics[width=1.0\linewidth]{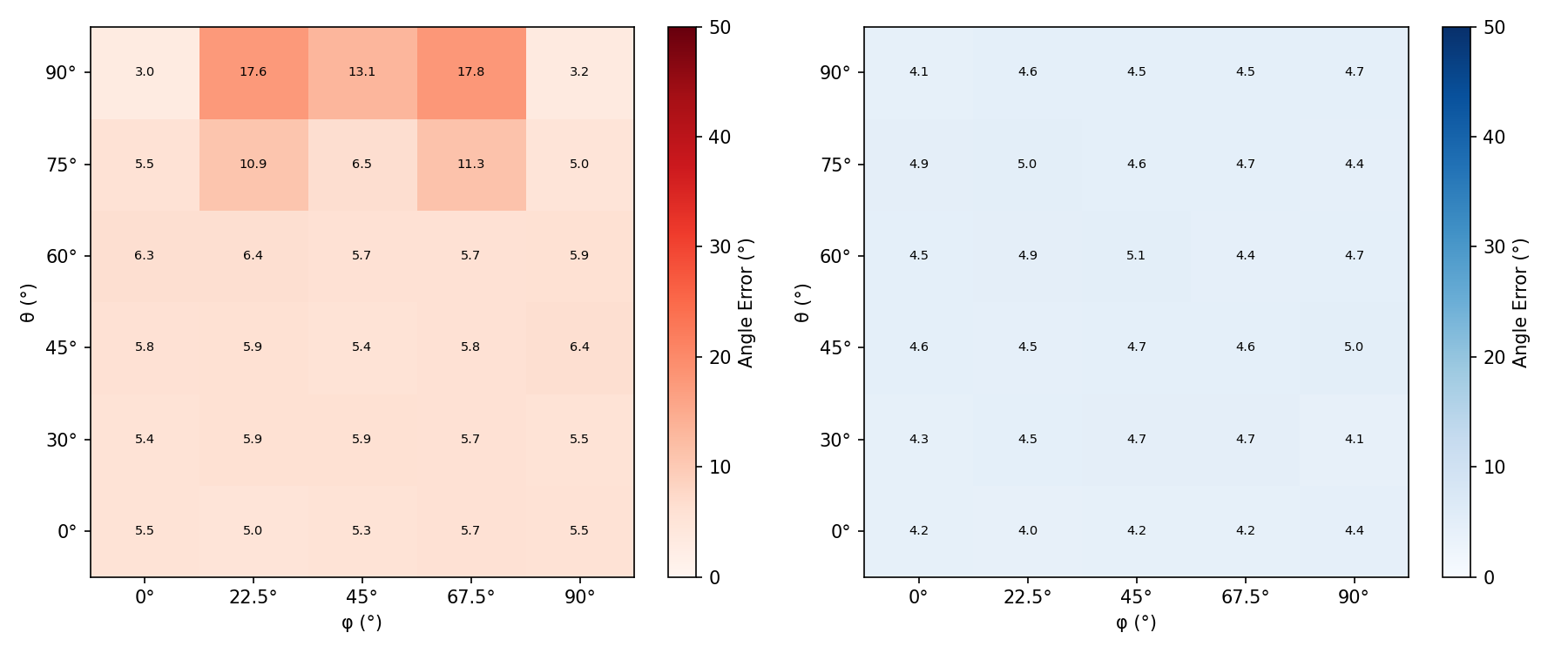}
    \caption{Angular resolution heatmap at 0.5 cm pitch for $\alpha = 20^\circ$.}
    \label{fig:angular_heatmap_20_5mm}
\end{figure}

\begin{figure}[H]
    \centering
    \includegraphics[width=1.0\linewidth]{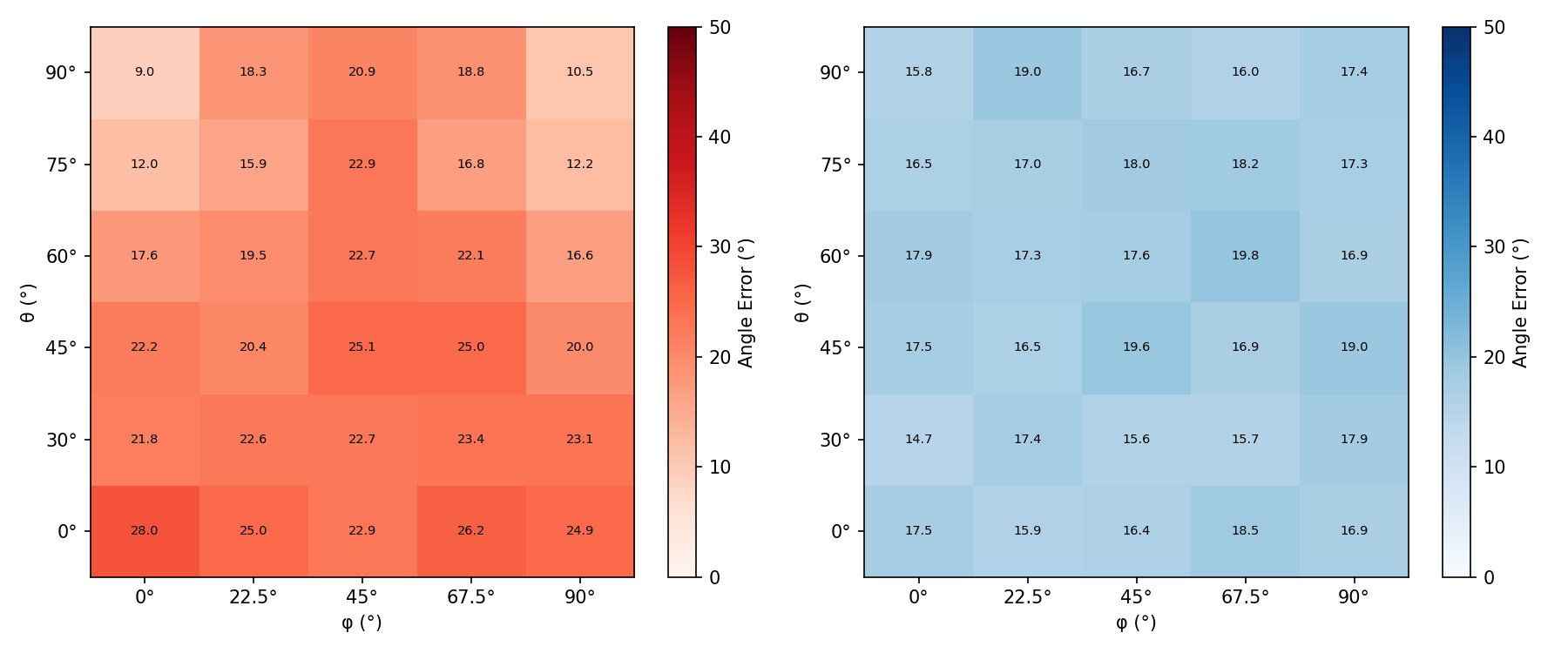}
    \caption{Angular resolution heatmap at 0.5 cm pitch for $\alpha = 40^\circ$ (shower-like).}
    \label{fig:angular_heatmap_40_5mm}
\end{figure}

Figures~\ref{fig:angular_heatmap_5_5mm}--\ref{fig:angular_heatmap_40_5mm} show angular resolution as a function of particle orientation $(\theta, \phi)$ for various cone opening angles at 0.5 cm pitch.

\subsection{Ghost Hit Heatmaps}

\begin{figure}[H]
    \centering
    \includegraphics[width=1.0\linewidth]{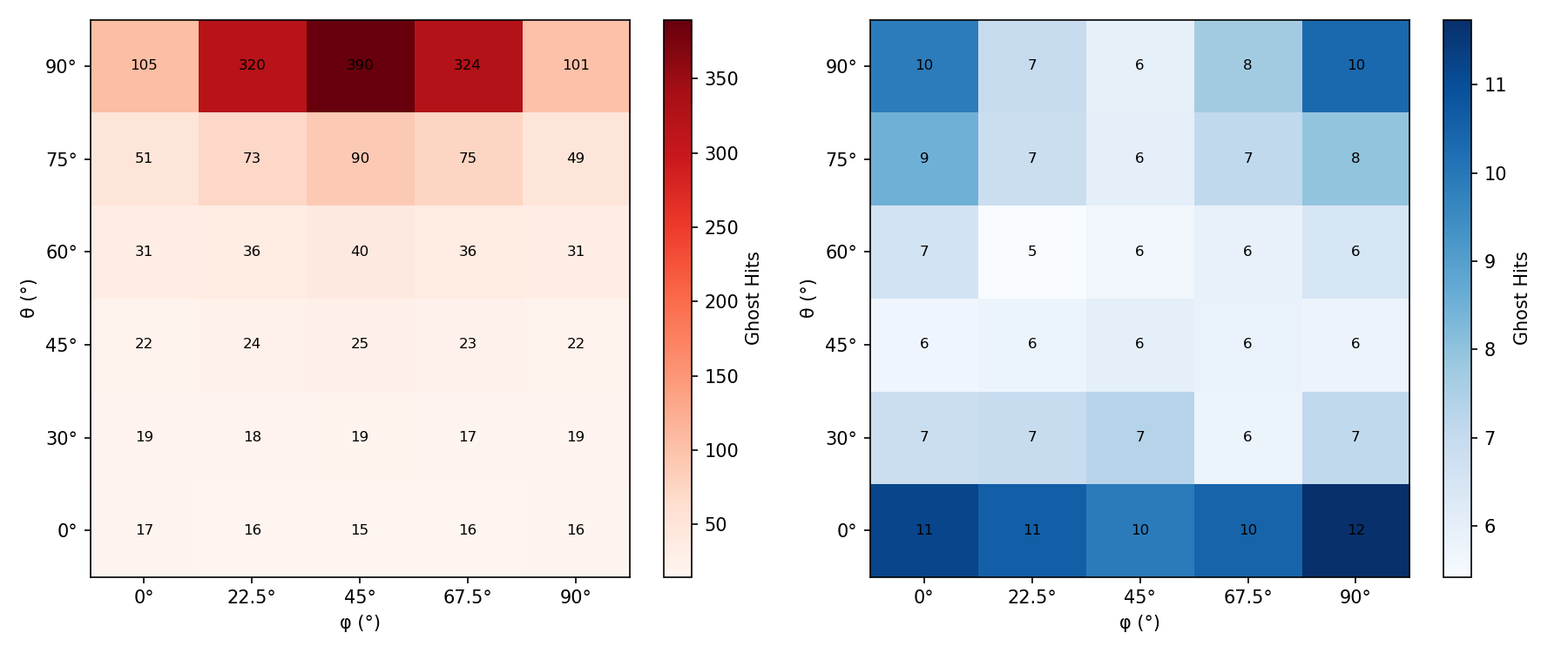}
    \caption{Ghost hit count heatmap at 0.5 cm pitch for $\alpha = 5^\circ$. Counts are identical to the 1.0 cm baseline, confirming scale independence.}
    \label{fig:ghost_heatmap_5_5mm}
\end{figure}

\begin{figure}[H]
    \centering
    \includegraphics[width=1.0\linewidth]{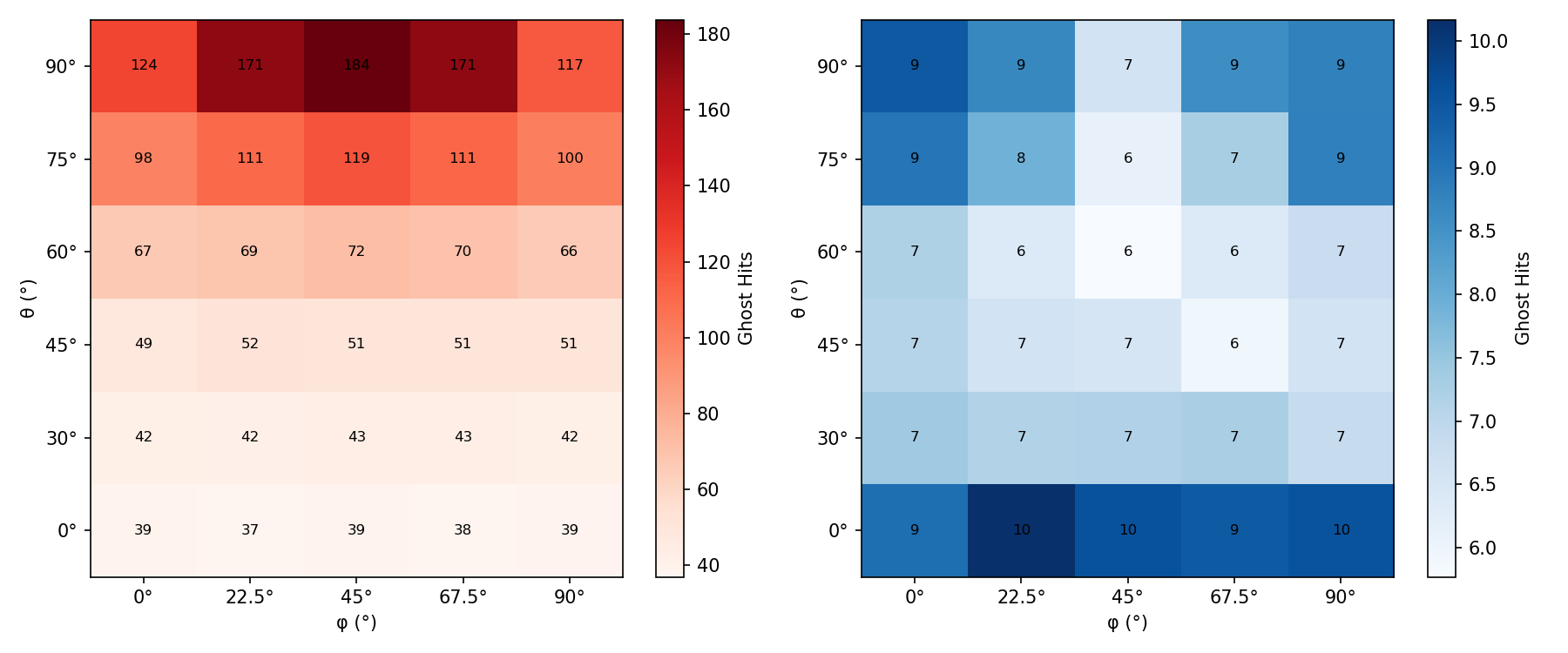}
    \caption{Ghost hit count heatmap at 0.5 cm pitch for $\alpha = 20^\circ$.}
    \label{fig:ghost_heatmap_20_5mm}
\end{figure}

\begin{figure}[H]
    \centering
    \includegraphics[width=1.0\linewidth]{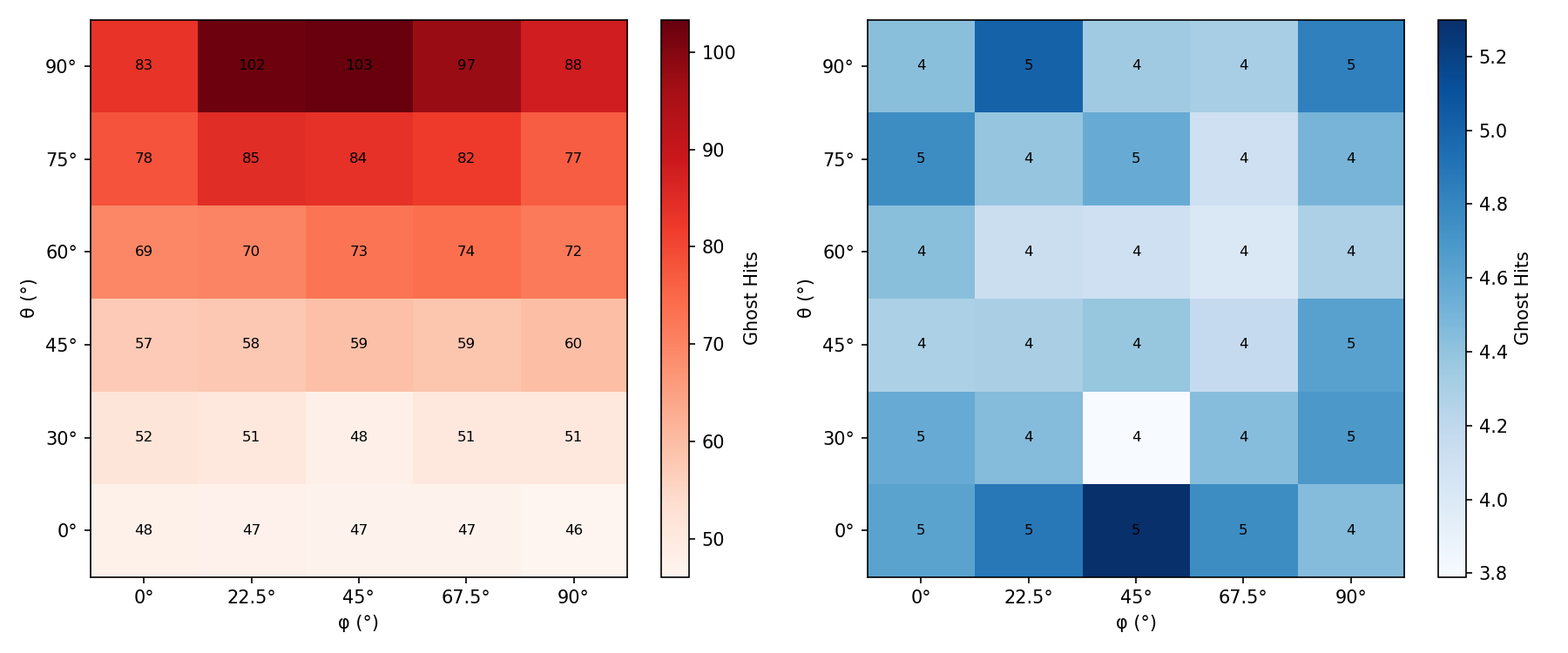}
    \caption{Ghost hit count heatmap at 0.5 cm pitch for $\alpha = 40^\circ$.}
    \label{fig:ghost_heatmap_40_5mm}
\end{figure}

Figures~\ref{fig:ghost_heatmap_5_5mm}--\ref{fig:ghost_heatmap_40_5mm} confirm that ghost hit counts are independent of pitch, as expected from the topological nature of the phenomenon.

\subsection{Single-Track Vertex Resolution}

\begin{figure}[H]
    \centering
    \includegraphics[width=1.0\linewidth]{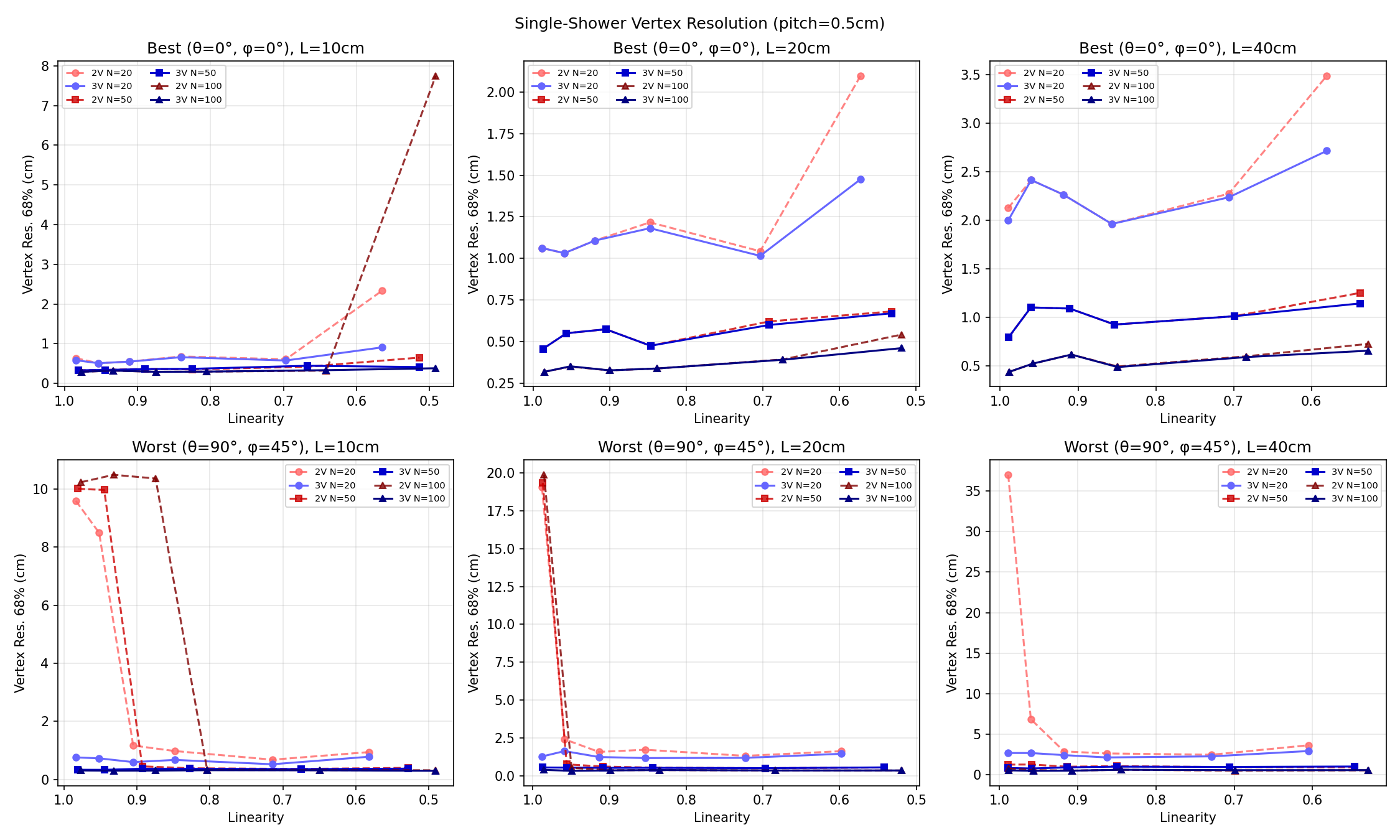}
    \caption{Single-track vertex resolution at 0.5 cm pitch for $N=50$ hits and $L = 10, 20, 40$ cm. Resolution improves to $\sigma_v \approx 0.15$ cm (compared to 0.29 cm at 1.0 cm pitch), consistent with $p/\sqrt{12}$ scaling.}
    \label{fig:vertex_L_5mm}
\end{figure}

Figure~\ref{fig:vertex_L_5mm} shows single-track vertex resolution as a function of track length at 0.5 cm pitch.

\subsection{Vertex Resolution Heatmaps}

\begin{figure}[H]
    \centering
    \includegraphics[width=1.0\linewidth]{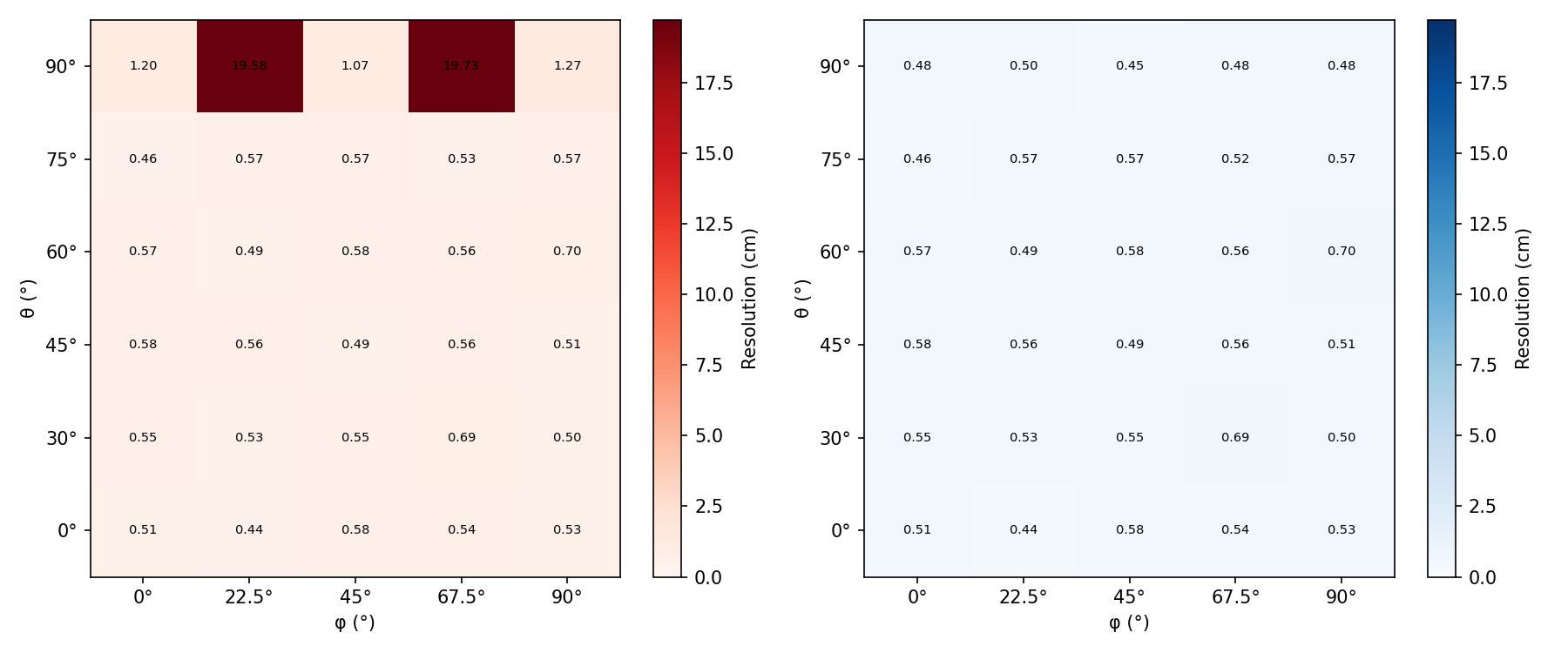}
    \caption{Vertex resolution heatmap at 0.5 cm pitch for $\alpha = 5^\circ$.}
    \label{fig:vertex_heatmap_5_5mm}
\end{figure}

\begin{figure}[H]
    \centering
    \includegraphics[width=1.0\linewidth]{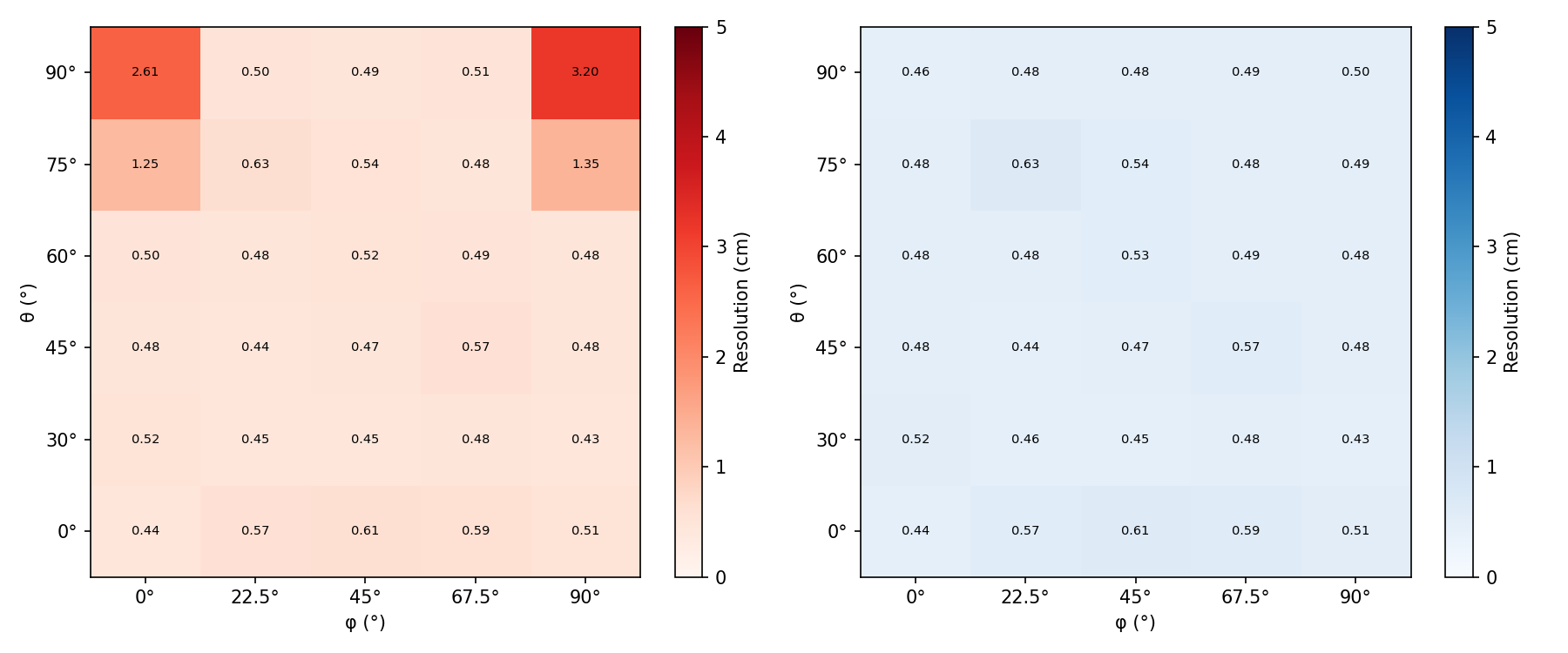}
    \caption{Vertex resolution heatmap at 0.5 cm pitch for $\alpha = 20^\circ$.}
    \label{fig:vertex_heatmap_20_5mm}
\end{figure}

\begin{figure}[H]
    \centering
    \includegraphics[width=1.0\linewidth]{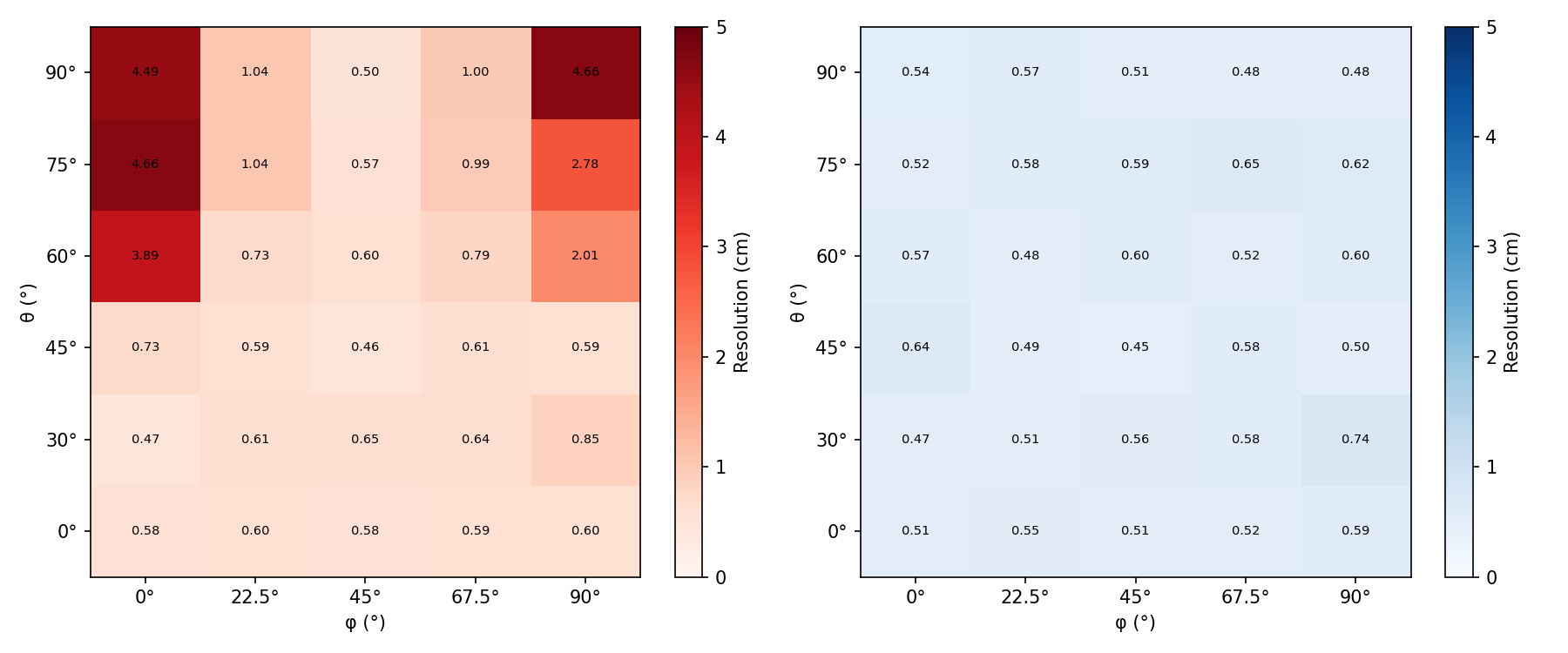}
    \caption{Vertex resolution heatmap at 0.5 cm pitch for $\alpha = 40^\circ$.}
    \label{fig:vertex_heatmap_40_5mm}
\end{figure}

Figures~\ref{fig:vertex_heatmap_5_5mm}--\ref{fig:vertex_heatmap_40_5mm} show vertex resolution as a function of orientation at 0.5 cm pitch.

A notable feature in Figure~\ref{fig:vertex_heatmap_5_5mm} is the sharp improvement in resolution (``dip'') at $\theta=90^\circ, \phi=45^\circ$ compared to neighboring angles. This effect arises from the interplay between the fine 0.5 cm pitch and the symmetric ghost distribution at $45^\circ$. The finer pitch allows the reconstruction algorithm to resolve the conical taper of the shower, which is obscured at 1.0 cm pitch due to coarse voxelization. At exactly $45^\circ$, the ghost pattern preserves this taper symmetry, enabling accurate vertex localization. At neighboring angles (e.g., $22.5^\circ$), the ghost distribution becomes asymmetric and blocky, disrupting the taper recognition and degrading resolution to levels similar to the 1.0 cm case.

\subsection{Energy-Dependent Performance}

\begin{figure}[H]
    \centering
    \includegraphics[width=1.0\linewidth]{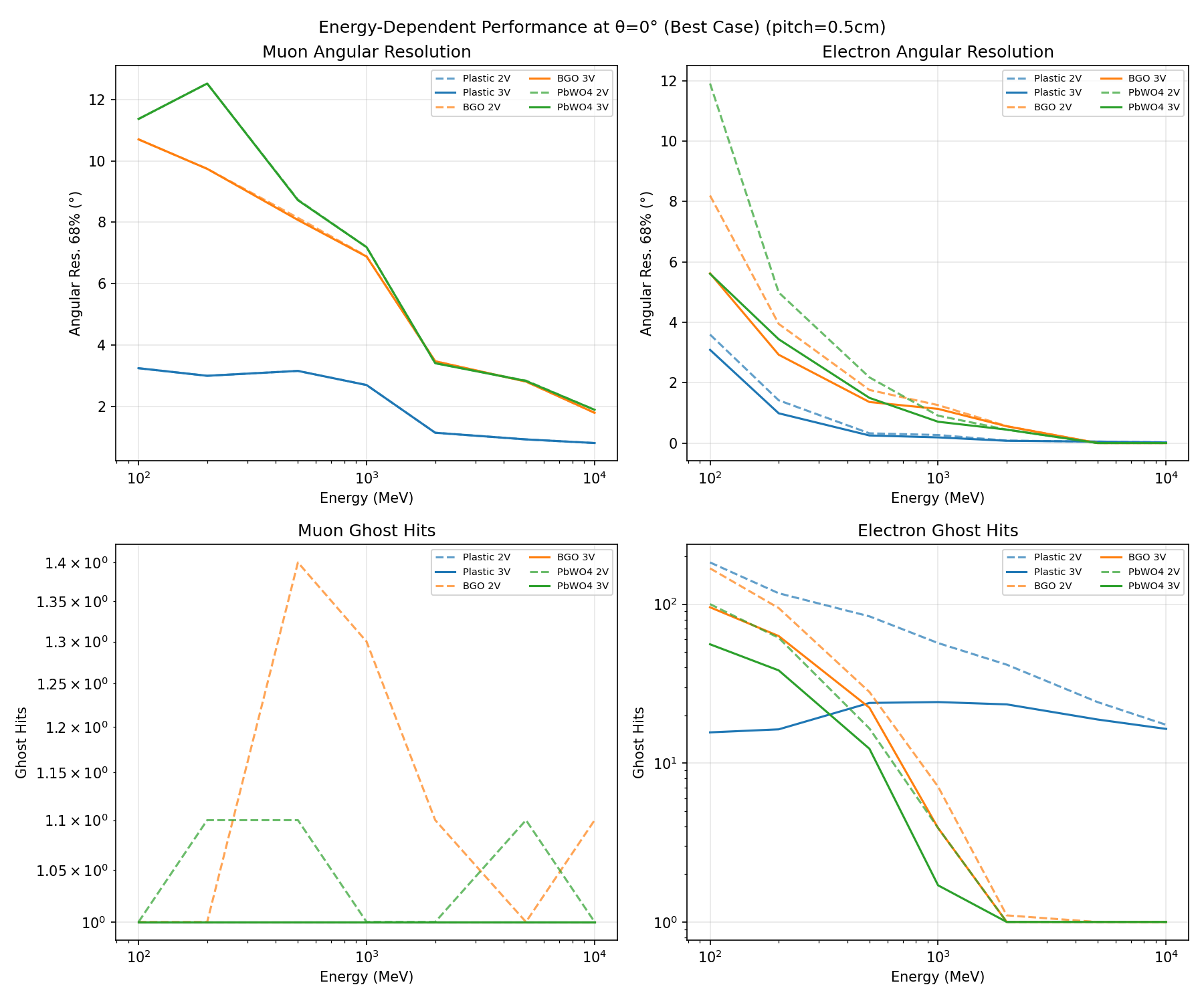}
    \caption{Energy-dependent performance at $\theta = 0^\circ$, pitch = 0.5 cm. Angular resolution improves by approximately 2$\times$ compared to 1.0 cm pitch across all energies and materials.}
    \label{fig:energy_theta0_5mm}
\end{figure}

\begin{figure}[H]
    \centering
    \includegraphics[width=1.0\linewidth]{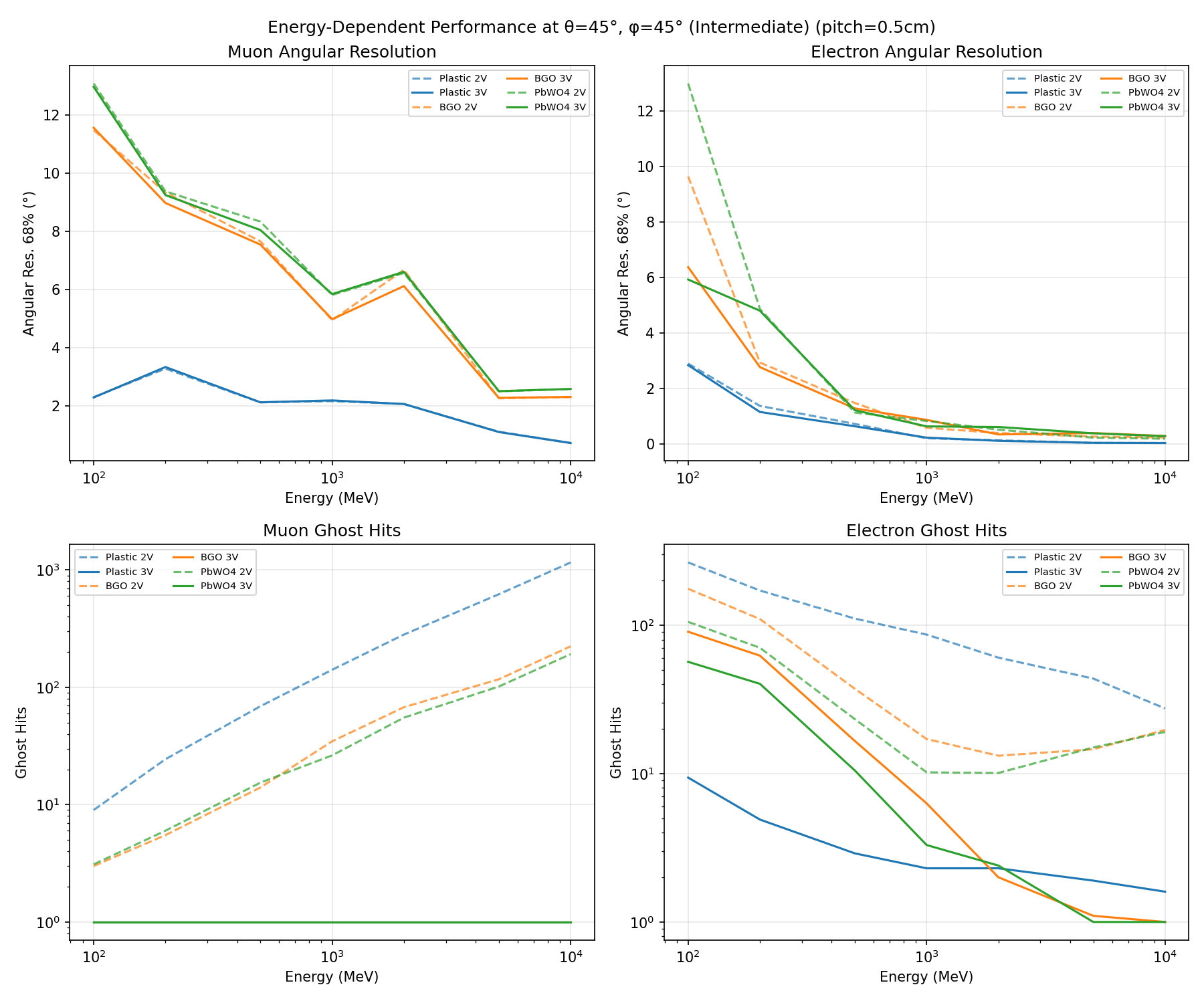}
    \caption{Energy-dependent performance at $\theta = 45^\circ$, $\phi = 45^\circ$, pitch = 0.5 cm.}
    \label{fig:energy_theta45_5mm}
\end{figure}

\begin{figure}[H]
    \centering
    \includegraphics[width=1.0\linewidth]{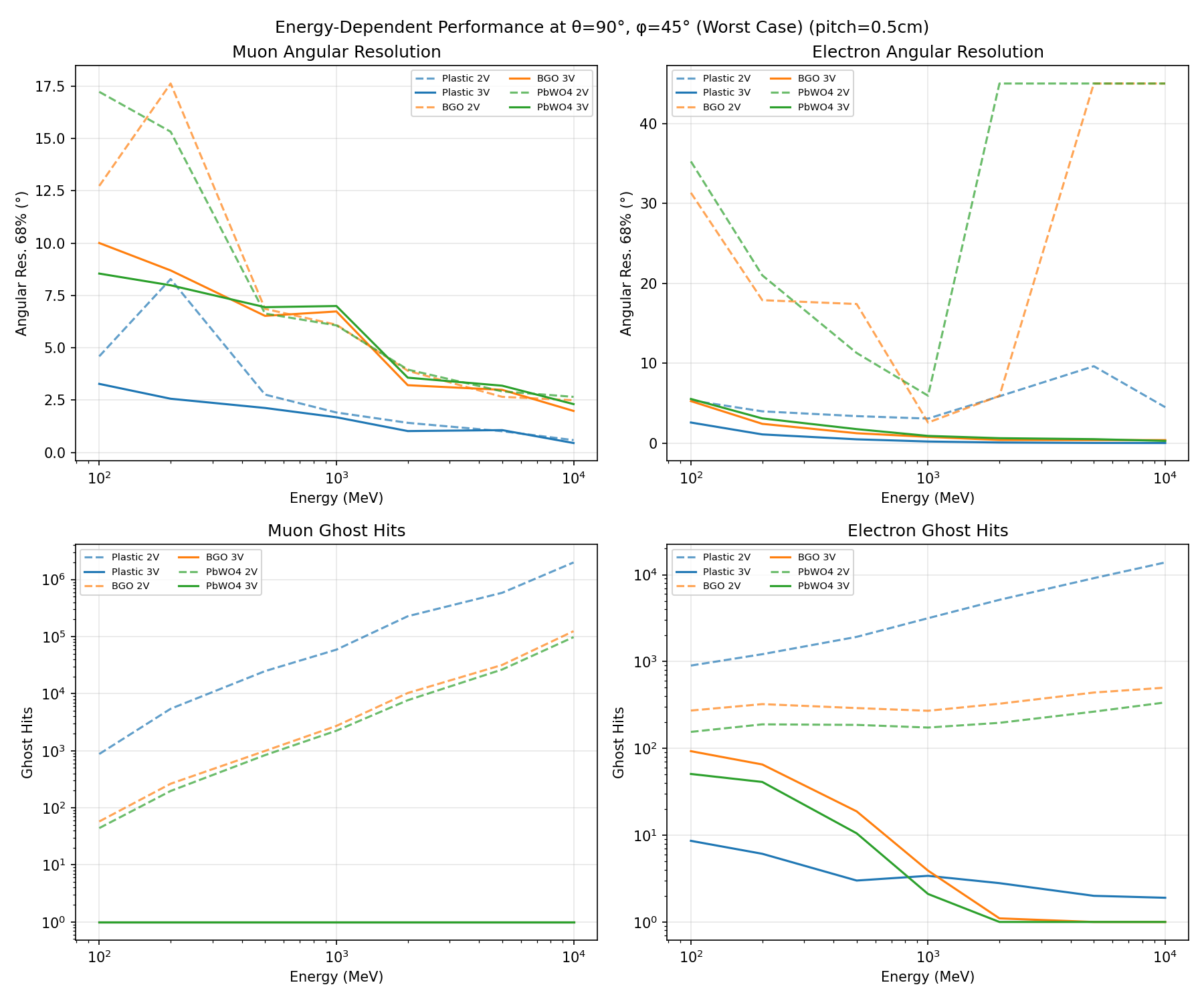}
    \caption{Energy-dependent performance at $\theta = 90^\circ$, $\phi = 45^\circ$, pitch = 0.5 cm. The dramatic 3V advantage at worst-case orientation persists: 2V resolution remains degraded to $\sim$20--30$^\circ$ at low energies, while 3V maintains $<$5$^\circ$ resolution.}
    \label{fig:energy_theta90_5mm}
\end{figure}

Figures~\ref{fig:energy_theta0_5mm}--\ref{fig:energy_theta90_5mm} show energy-dependent angular resolution and ghost hits at 0.5 cm pitch for three orientations, using physics-derived track parameters.

\subsection{Summary and Scaling Validation}

Table~\ref{tab:pitch_scaling} summarizes the resolution scaling from 1.0 cm to 0.5 cm pitch.

\begin{table}[h]
\centering
\caption{Resolution scaling from 1.0 cm to 0.5 cm pitch.}
\label{tab:pitch_scaling}
\begin{tabular}{lccc}
\hline
\textbf{Metric} & \textbf{1.0 cm} & \textbf{0.5 cm} & \textbf{Ratio} \\
\hline
Angular resolution (3V, $\alpha=5^\circ$) & $\sim$2$^\circ$ & $\sim$1$^\circ$ & 2.0 \\
Vertex resolution (single track) & 0.29 cm & 0.15 cm & 1.9 \\
Two-cluster vertex (3V, $\alpha=5^\circ$) & $\sim$2 cm & $\sim$1 cm & 2.0 \\
Ghost hit count ($N=50$, $\alpha=20^\circ$) & $\sim$100 & $\sim$100 & 1.0 \\
\hline
\end{tabular}
\end{table}

Key observations:
\begin{itemize}
    \item \textbf{Linear scaling}: Angular and vertex resolution improve by a factor of $\sim$2, consistent with the pitch ratio.
    \item \textbf{Ghost independence}: Ghost hit counts are identical, confirming the topological nature of the 3V advantage.
    \item \textbf{3V advantage preserved}: The relative performance gain of 3V over 2V (factor of 2--10$\times$ depending on topology and orientation) is maintained across pitch sizes.
    \item \textbf{Practical implications}: At 0.5 cm pitch, 3V achieves sub-degree angular resolution and sub-centimeter vertex resolution across most orientations and energies, enabling precision measurements in neutrino oscillation experiments and rare decay searches.
\end{itemize}

\end{document}